\pdfoutput=1
\documentclass[11pt,twoside,a4paper,cmspaper,final,collab]{cms-tdr}
\def\svnVersion{fad16f5}\def\svnDate{2024/08/15}\def\cmsCernNoTag{CERN-EP-2024-010}\def\cmsCernDate{\today}\def\cmsMessage{Published in Physical Review Letters as \href{http://dx.doi.org/10.1103/PhysRevLett.133.071903}{\doi{10.1103/PhysRevLett.133.071903}.}}
\begin{document}\cmsNoteHeader{SMP-22-015}

\ifthenelse{\boolean{cms@external}}
{\providecommand{\suppMaterial}{the supplemental material~\cite{suppMatBib}}}
{\providecommand{\suppMaterial}{Appendix~\ref{app:suppMat}\nocite{suppMatBib2}}}
\newlength\cmsFigWidth
\ifthenelse{\boolean{cms@external}}{\setlength\cmsFigWidth{0.35\textwidth}}{\setlength\cmsFigWidth{0.5\textwidth}}
\newlength\cmsTabSkip\setlength\cmsTabSkip{1.2ex}
\newcommand{\asZ}{\ensuremath{\alpS (m_\PZ)}\xspace}
\newcommand{\dR}{\ensuremath{\Delta R}\xspace}
\newcommand{\ptjet}{\ensuremath{\pt^{\text{jet}}}\xspace}
\newcommand{\DIRE} {\textsc{dire}\xspace}
\newcommand{\xL} {\ensuremath{x_\text{L}}\xspace}
\newcommand{\VINCIA} {\textsc{vincia}\xspace}
\newcommand{\NNLL} {\ensuremath{\text{NLO}+\text{NNLL}_\text{approx}}\xspace}

\cmsNoteHeader{SMP-22-015}

\title{Measurement of energy correlators inside jets and determination of the strong coupling \texorpdfstring{\asZ}{alphaS(mZ)}}
\date{\today}

\abstract{Energy correlators that describe energy-weighted distances between two or three particles in a hadronic jet are measured using an event sample of $\sqrt{s} =13\TeV$ proton-proton collisions collected by the CMS experiment and corresponding to an integrated luminosity of 36.3\fbinv. The measured distributions are consistent with the trends in the simulation that reveal two key features of the strong interaction: confinement and asymptotic freedom. By comparing the ratio of the measured three- and two-particle energy correlator distributions with theoretical calculations that resum collinear emissions at approximate next-to-next-to-leading logarithmic accuracy matched to a next-to-leading order calculation, the strong coupling is determined at the \PZ boson mass: $\asZ =0.1229^{+0.0040}_{-0.0050}$, the most precise \asZ value obtained using jet substructure observables.}

\hypersetup{%
pdfauthor={CMS Collaboration},%
pdftitle={Measurement of energy correlators inside jets and determination of the strong coupling alphaS(mZ)},%
pdfsubject={CMS},%
pdfkeywords={CMS, jet substructure, strong coupling, E2C, E3C}}

\maketitle

In high energy proton-proton ($\Pp\Pp$) collisions, 
quarks and gluons undergo cascades of radiation and hadronization processes, 
resulting in the production of hadrons that can cluster in jets. 
Collisions at the LHC have produced an unprecedented number of events containing jets,
spanning a broad range of energies. 
Studies of jets provide information on the strong interaction, 
described by quantum chromodynamics (QCD)~\cite{Gross:2022hyw}. 
Jet substructure has been widely studied at the LHC~\cite{ATLAS:2020bbn,ATLAS:2017pgl,ATLAS:2019mgf,
CMS:2021iwu,CMS:2017qlm,CMS:2018vzn,
ALICE:2021njq,ALargeIonColliderExperiment:2021mqf,ALICE:2021aqk,LHCb:2019qoc}.
Comparing high precision measurements of jet substructure observables
with Monte Carlo (MC) simulations and perturbative QCD (pQCD) calculations
improves our understanding of QCD.

New observables have been recently proposed to describe multiparticle
energy correlations within jets~\cite{Chen:2020vvp}. 
They evolved from the event shape observable energy-energy correlator (EEC)~\cite{Basham:1978bw}, 
which was extensively studied in $\EE$ experiments~\cite{PLUTO:1981gcc,L3:1992btq,OPAL:1991uui,DELPHI:1992qrr}.
The new observables study correlations inside jets, 
therefore being more sensitive than EEC in the small angle limit~\cite{Hofman:2008ar}. 
They are derived from energy-flow operators, 
are collinear- and infrared-safe~\cite{Salam:2010nqg}, and
can be formulated with correlation functions of 
light-ray operators~\cite{Sveshnikov:1995vi,Belitsky:2013xxa,Belitsky:2013bja,Kologlu:2019mfz}
widely used in conformal field theory.
Analytical calculations with all-orders resummation 
at next-to-leading logarithmic (NLL)~\cite{Dixon:2019uzg,Lee:2022ige} 
and approximate next-to-NLL ($\text{NNLL}_\text{approx}$)~\cite{Chen:2023zlx} accuracy in pQCD, 
matched to next-to-leading order (NLO) pQCD calculations, 
denoted as \NNLL, 
can be achieved for these observables. 
This allows comparisons between theory and data with 5\% accuracy~\cite{Chen:2023zlx}. 
We focus on the two- and three-particle energy correlators 
(E2C and E3C, respectively),
for which there are predictions with \NNLL precision~\cite{Chen:2023zlx}. 

Studies of jet substructure can be used to determine 
the strong coupling \alpS~\cite{Proceedings:2018jsb,Salam:2017qdl,Moch:2014tta,LeBlanc:2022bwd}. 
Among the interaction couplings of the standard model, \alpS is the least precisely known~\cite{ParticleDataGroup:2022pth}. 
The world average \alpS at the \PZ boson mass is $\asZ = 0.1180$, 
with a 0.8\% precision~\cite{ParticleDataGroup:2022pth}. 
Methods that mainly rely on fixed-order calculations
provide \alpS extractions with 1--3\% precision at the 
LHC~\cite{ATLAS:2023lhg,CMS:2021yzl,ATLAS:2023tgo,CMS:2018fks,dEnterria:2019aat,Klijnsma:2017eqp};
jet substructure is more sensitive to collinear physics effects, 
allowing for measurements in complementary phase space regions. 
In these regions, studies based on event shapes in 
$\EE$ collisions at LEP~\cite{Abbate:2010xh,Gehrmann:2012sc,Hoang:2015hka} 
obtained a value of \alpS about 5\% lower than the world average, 
emphasizing the importance of new studies. 
It has been estimated that a value of \alpS with 10\% precision 
could be achieved using the soft-drop groomed jet mass~\cite{Proceedings:2018jsb}. 
There are two major difficulties in reaching a high-precision \alpS extraction with this observable:
its dependence on the relative fraction of quark- and gluon-initiated jets, 
since a larger gluon jet fraction can be misinterpreted as a larger \alpS value;
and the large uncertainty in the nonperturbative corrections.
A jet substructure measurement using heavy-flavor jets 
provided an \alpS value with $\approx$\,10\% precision~\cite{CMS:2018ypj}. 
Other approaches have been proposed to tackle the parton-flavor dependence~\cite{LeBlanc:2022bwd}. 
An approach based on the ratio of multiparticle energy correlators
was proposed to solve both problems and improve 
the precision of jet substructure \alpS determinations~\cite{Chen:2020vvp}.

Unlike observables such as the jet charge, which provide a single value for each jet, 
E2C and E3C provide distributions because each jet has many particles and, hence,
many pairs and triplets.
These distributions depend on \xL, 
defined as the angular distance between the two particles of the E2C pair, 
$\sqrt{\smash[b]{(\Delta\eta_{i,j})^2+(\Delta\phi_{i,j})^2}}$, 
where $\Delta\eta$ and $\Delta\phi$ are the distances 
in pseudorapidity and azimuthal angle of the particles, respectively,
or as the largest angular distance among the three pair combinations of an E3C triplet.
A jet with $n$ particles contributes $n^2$ entries to the E2C \xL distribution,
each having a weight $E_{i} E_{j} / E^2$, 
the product of the fractional energy of the jet carried by each of the particles.
The entry with $j=i$ adds to the $\xL = 0$ bin, 
which contributes to the overall normalization of the distribution.
Similarly, a jet contributes $n^3$ entries to the E3C \xL distribution,
with weights $E_{i} E_{j} E_{k} / E^3$,
where the indices represent the three particles in the triplet.
The fractional energy weights ensure that E2C and E3C
are insensitive to soft radiation, 
without the need for jet grooming techniques~\cite{Chen:2020vvp},
and that their distributions are normalized to the number of jets in the event sample. 
Additionally, taking the ratio of two energy correlators further suppresses nonperturbative effects.

The distinctive shape of the \xL-differential E2C and E3C correlators
provides information on the time scale of hadron formation, 
still an open question in QCD~\cite{Gross:2022hyw}. 
The different \xL regions probe the dynamics of jet formation,
governed by the DGLAP
evolution equations~\cite{Gribov:1972ri,Dokshitzer:1977sg,Marchesini:1994wr,Altarelli:1977zs}.
Most jet substructure observables are dominated by terms proportional to
the coupling \alpS, to a large logarithm due to collinear divergences $\ln \xL$, 
and to a color factor $C_i$~\cite{Proceedings:2018jsb}.
The quark-gluon composition, essential to extract \alpS, 
can be calculated from parton distribution functions (PDFs), 
at the price of introducing large uncertainties~\cite{Proceedings:2018jsb}. 
The PDF dependence is largely suppressed in the E3C/E2C ratio,
which remains sensitive to \alpS because E3C and E2C have different dependences on $\alpS$, $C_i$ 
and its higher order expansion terms.
The calculation at leading logarithmic accuracy implies that the ratio is 
approximately linear in $\alpS \ln \xL$~\cite{Chen:2020vvp},
so that measuring its slope with respect to $\ln \xL$, 
as a function of the jet momentum transverse to the $\Pp\Pp$ collision axis (\pt),
gives the dependence of \alpS on the jet energy.

In this Letter we report the E2C and E3C distributions versus \xL
and the \asZ value obtained by comparing their ratio 
with theoretical predictions at \NNLL accuracy~\cite{Chen:2023zlx}. 
This is the first measurement of E2C at a hadron collider, 
probing a much higher energy scale than at the $\EE$ experiments, 
and the first ever measurement of E3C.
The analysis is based on an event sample of $\Pp\Pp$ collisions at $\sqrt{s} = 13$\TeV
collected by CMS in 2016 
and corresponding to an integrated luminosity of 36.3\fbinv~\cite{CMS-LUM-17-003}.
The CMS detector has been designed~\cite{CMS:2008xjf} 
to trigger on~\cite{CMS:2020cmk,CMS:2016ngn} 
and identify electrons, muons, photons, and 
hadrons~\cite{CMS:2020uim,CMS:2018rym,CMS:2014pgm,CMS:2019ctu}. 
The particles are reconstructed~\cite{CMS:2017yfk,CMS:2018jrd,CMS:2016lmd} 
from information provided 
by a silicon tracker, an electromagnetic calorimeter, and a hadron calorimeter, 
placed in a 3.8\unit{T} superconducting solenoid, 
and by muon detectors placed outside.

The hadronic jets are clustered with the anti-\kt algorithm~\cite{Cacciari:2008gp, Cacciari:2011ma}
with a resolution parameter of 0.4, 
a value suited for parton shower (PS) studies at relatively small angles~\cite{Chen:2023zlx,Lee:2022ige}. 
The jet four-momentum is determined as the vectorial sum of all particle four-momenta in the jet. 
We select the dijet process to measure E2C and E3C because of its large cross section and wide energy range. 
Background processes beyond those produced by the strong interaction 
have a negligible contribution~\cite{CMS:2021yzl}, 
allowing for a direct comparison between data and predictions. 
A series of eight triggers are used, each requiring a leading jet with increasing \pt thresholds, 
the lowest one being 60\GeV.
The events must contain at least two jets, with $\pt > 97\GeV$ and $\abs{\eta} < 2.1$. 
The $\eta$ selection ensures that the jet constituents are reconstructed with good spatial and momentum resolutions. 
In addition, the two leading jets are required to be back-to-back in the transverse plane, 
with $\abs{\Delta \phi } > 2$. 
Jets with lower \pt are not considered because the corresponding pQCD calculations 
can only be performed in a limited \xL range, so that the different stages in jet formation cannot be reliably identified. 
To study the E2C and E3C energy dependence, 
the two leading jets in each event are binned in eight \pt bins, in the 97--1784\GeV range. 
The boundaries are optimized so that the corresponding trigger has an efficiency exceeding 99.5\%~\cite{CMS:2021yzl}. 
Weights corresponding to the trigger prescales are applied~\cite{CMS:2020cmk}. 
All the neutral and charged particles with $\pt > 1\GeV$ within the jets are used to compute the energy correlators. 
Each jet contributes multiple entries to the E2C and E3C distributions
and each event provides two leading jets, 
causing statistical correlations across \xL and \pt bins,
amounting to 40\% on average. 
Covariance matrices are calculated for the two distributions. 
Since the analysis is not statistically limited, 
we measure the E2C and E3C distributions from two independent data samples, 
built by splitting the data in two halves,
to avoid keeping track of the correlations between the two variables,
which would lead to larger matrices and computational challenges.

In order to obtain the E2C and E3C distributions at particle level, 
we unfold the measured distributions for the detector response,
using the iterative D'Agostini unfolding method~\cite{DAgostini}, 
implemented in \textsc{RooUnfold}~\cite{Adye:2011gm,Brenner:2019lmf}.
We stop the iterations once the $p$-value between consecutive unfolded distributions exceeds 0.05,
meaning that the distributions become statistically consistent.
The reconstructed jets and particles are considered matched to the generated ones 
if their relative angular distance is smaller than 0.2 and 0.05, respectively. 
If multiple matches are found, the one with the closest \pt is chosen. 
The jet matching efficiency is above 99\%. 
The fractions of correctly matched charged particles, photons, and neutral particles 
are 80--84\%, 54--70\%, and 20--35\%, respectively, 
corresponding to 88--95\%, 80--97\%, and 40--92\% of the total energy of these particle types.
The unmatched objects are considered to be misreconstructed and their contribution is subtracted from the data.
After unfolding, bin-by-bin inefficiency corrections are applied, derived from the particle-level unmatched objects.
Given the correlations between the variables, the unfolding is performed in three dimensions: 
\xL, \pt, and the energy weight.
Simulations are used to derive the response matrices.

Four samples of simulated events are used to model hadronization effects and to evaluate the detector response.
They are generated with \PYTHIA{8.240}~\cite{Sjostrand:2014zea}, 
\HERWIG{7.1.4}~\cite{Bahr:2008pv,Bellm:2015jjp}, 
and \MGvATNLO\cite{Alwall:2014hca,Alwall:2007fs}, 
the latter being interfaced with \PYTHIA{8} or \HERWIG{7}. 
Dijet events are generated at leading order (LO) in pQCD, with NNPDF3.1 PDFs~\cite{Ball:2014uwa}, 
and interfaced with \GEANTfour~\cite{Agostinelli:2002hh}.
In the \MGvATNLO samples,
final states with up to four partons are generated at tree level, 
and the MLM jet matching scheme is applied~\cite{Alwall:2007fs}. 
Simulated pileup interactions are included, 
normalized to a total inelastic cross section of 69.2\unit{mb}\cite{CMS:2020ebo}. 
The unfolding uses the \PYTHIA{8} event sample, which is the largest one.
Other event samples are generated at the particle level to provide predictions 
that can be directly compared with the (unfolded) measured distributions.
They reflect several alternative PS and hadronization models. 
More details of the simulated samples can be found in \suppMaterial. 

The energy scale uncertainty of the jet constituents, 
3\% for photons, 5\% for neutral particles, and 1\% for charged particles, 
affects the energy of these particles and, hence, the jet \pt. 
The track reconstruction efficiency uncertainty, 3\%, reflects the mismodeling of the efficiency 
to reconstruct charged-particle tracks in the dense core of the jets~\cite{CMS:2016lmd}. 
The MC event generators differ in the PS and hadronization modeling, 
in the tuning of parameters, and in the fixed-order matrix calculation for the hard scattering.
These differences are not covered by the renormalization scale uncertainty
nor by varying the underlying event (UE) in \PYTHIA{8}. 
Therefore, we use the differences between the results obtained with the baseline \PYTHIA{8} MC 
and the other samples to evaluate one-sided uncertainties in the MC modeling.
The (unfolded) particle-level data distributions are recomputed with the responses
corresponding to the variations mentioned above, to establish their uncertainty.
The variations in MC modeling contribute the largest source of uncertainty,
2--10\% depending on \xL and \pt, 
followed by the neutral particle energy scale, which contributes an uncertainty of 1--2\%.

\begin{figure*}[ht]
\centering
\includegraphics[width=\linewidth]{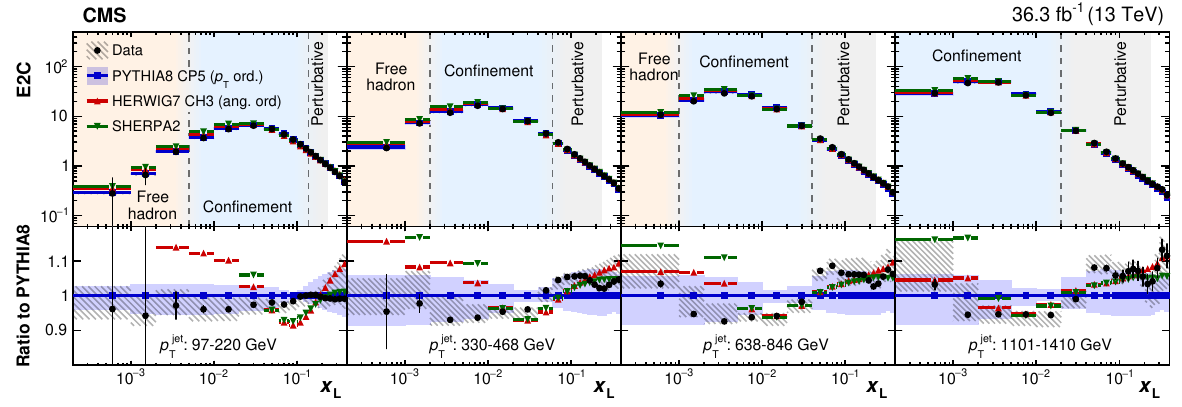}
\caption{Measured (unfolded) and simulated E2C \xL distributions, 
in four \pt bins.
The lower panels show the ratios to the \PYTHIA{8} reference. 
The data statistical (bars) and systematic (boxes) uncertainties are also shown, as is
the \PYTHIA{8} uncertainty (blue band).}
\label{fig:dataUnfoldE2C}
\end{figure*}

The measured (unfolded) E2C distributions 
are shown in Fig.~\ref{fig:dataUnfoldE2C}, 
in four illustrative jet \pt ranges,
together with the predictions of three models:
\PYTHIA{8} (with the CP5 tune and \pt-ordered showers), 
\HERWIG{7} (with the CH3 tune and angular-ordered showers), 
and \SHERPA{2} (with default settings).
The E2C distributions in the other four jet \pt ranges,
the E3C distributions, and 
comparisons with all the considered MC models
are collected in \suppMaterial, while 
tabulated results are provided in the HEPData record for this analysis~\cite{hepdata}.
The lower panels show the ratios between 
the data and the \PYTHIA{8} reference.
The uncertainty in the \PYTHIA{8} MC prediction (blue band)
reflects the following sources.
The uncertainties from the missing higher orders in the hard scattering computation 
and from the PS modeling are obtained by independently varying 
the renormalization and factorization scales by factors of 0.5 and 2, 
keeping their ratio between 0.5 and 2.
The PDF uncertainty is evaluated using an envelope of 100 PDF sets,
corresponding to the variations of the uncertainty eigenvectors of the default set. 
The uncertainty in the infrared approximation of the \PYTHIA{8} PS splitting kernels 
is evaluated by varying the coefficient of the nonsingular term by $\pm 2$~\cite{Mrenna:2016sih}. 
The uncertainty in the UE model is evaluated by using
the Monash tune~\cite{Skands:2014pea}.
The measured and \PYTHIA{8} distributions show good agreement,
given the uncertainties.
The lower panels of Fig.~\ref{fig:dataUnfoldE2C} also show the ratios between the
\HERWIG{7} and \SHERPA{2} MC distributions and the \PYTHIA{8} reference,
to illustrate the level of variation that exists among models.

Since the MC models provide a reasonably good description of 
the measured \xL dependence of E2C,
we can discuss it in terms of three phases in the evolution of the produced jets. 
The momentum exchange between two particles is proportional to $\pt \xL$~\cite{Chen:2020vvp}, 
so that $\xL$ reflects the energy scale of the interaction.
In the large \xL region, dominated by wide-angle splittings 
from the emission of additional partons during the PS stage, 
we see that E2C decreases as \xL increases, 
as predicted by pQCD~\cite{Lee:2022ige}. 
The small \xL region, where we have the opposite trend, 
reflects a phase dominated by noninteracting hadrons.
The intermediate \xL region corresponds to a transition phase, 
where the partons get confined in the final hadrons.
To determine the \xL boundaries,
shown as dashed vertical lines in Fig.~\ref{fig:dataUnfoldE2C},
we fit the \xL distributions in each jet \pt range and 
identify the regions that follow the quantitative scaling predictions:
in the free-hadron region, the E2C and E3C particle-level distributions 
are expected to increase with \xL as $\exp(2\ln \xL)$~\cite{Komiske:2022enw}; 
in the perturbative region, the E3C/E2C ratio of parton-level distributions
is expected to increase with \xL as $\ln \xL$~\cite{Chen:2023zlx}, 
with small differences at the hadron level. 
The fits of the parton- and hadron-level distributions are made using the simulated trends, 
which describe well the shapes of the measured distributions.
As the jet \pt increases, the boundaries shift towards smaller \xL,
so that the energy scale at which the transition occurs, 
$Q = \mathrm{a} \pt \xL$~\cite{Chen:2020vvp}, remains the same.
The constant $a$ is unknown but 
the boundaries derived from simulation suggest that $Q/a \approx 20\GeV$ 
for the transition between the perturbative and confinement regions 
and $\approx$\,0.8\GeV for the transition between the confinement and free-hadron regions. 
The boundaries are sensitive to \alpS.
We only compare data and pQCD where the calculations are reliable.

\begin{figure}[ht]
\centering
\includegraphics[width=0.49\textwidth]{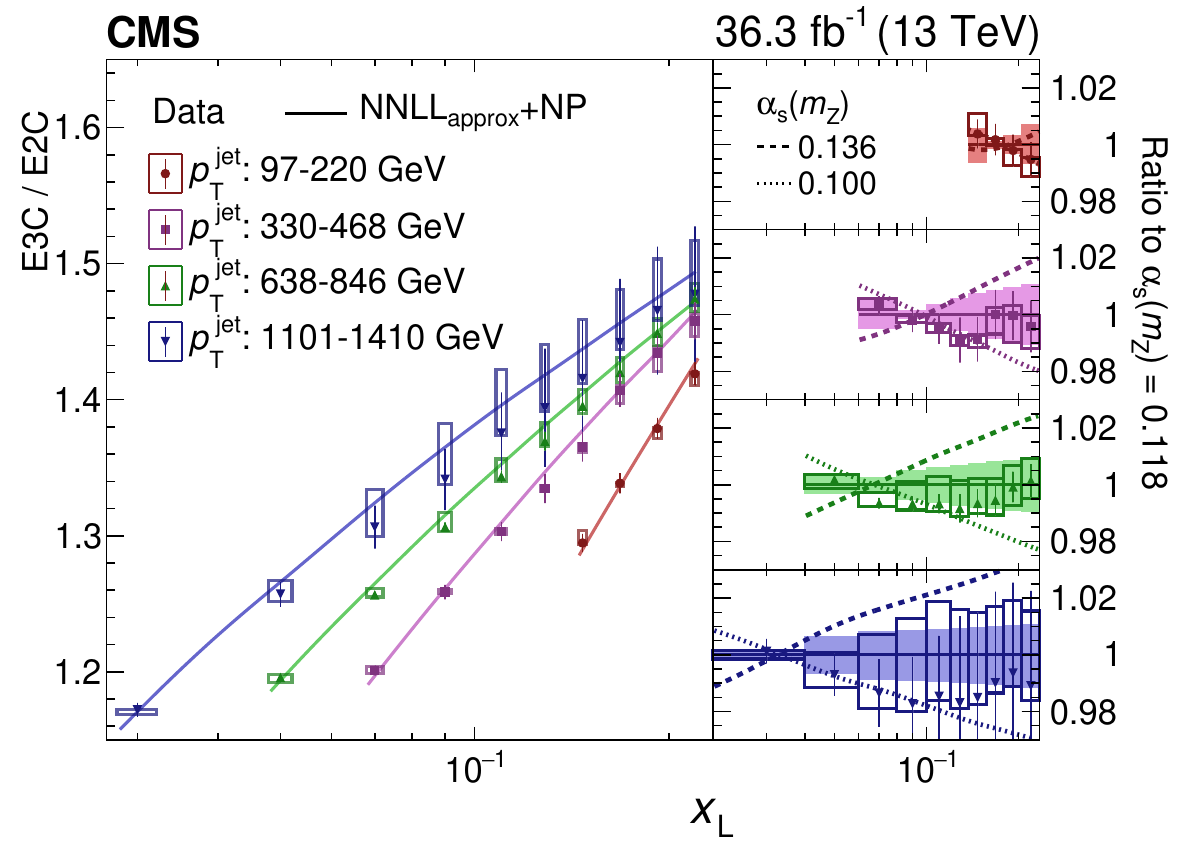}
\caption{Measured E3C/E2C ratio (left) and their ratio to predictions (right) 
in the perturbative \xL region and four jet \pt bins.
The \NNLL predictions~\cite{Chen:2023zlx} 
are corrected to hadron-level and normalized to the data. 
The statistical and experimental systematic uncertainties 
are shown with bars and boxes, respectively.}
\label{fig:dataUnfoldRatio}
\end{figure}

\begin{figure}[ht!]
\centering
\includegraphics[width=0.49\textwidth]{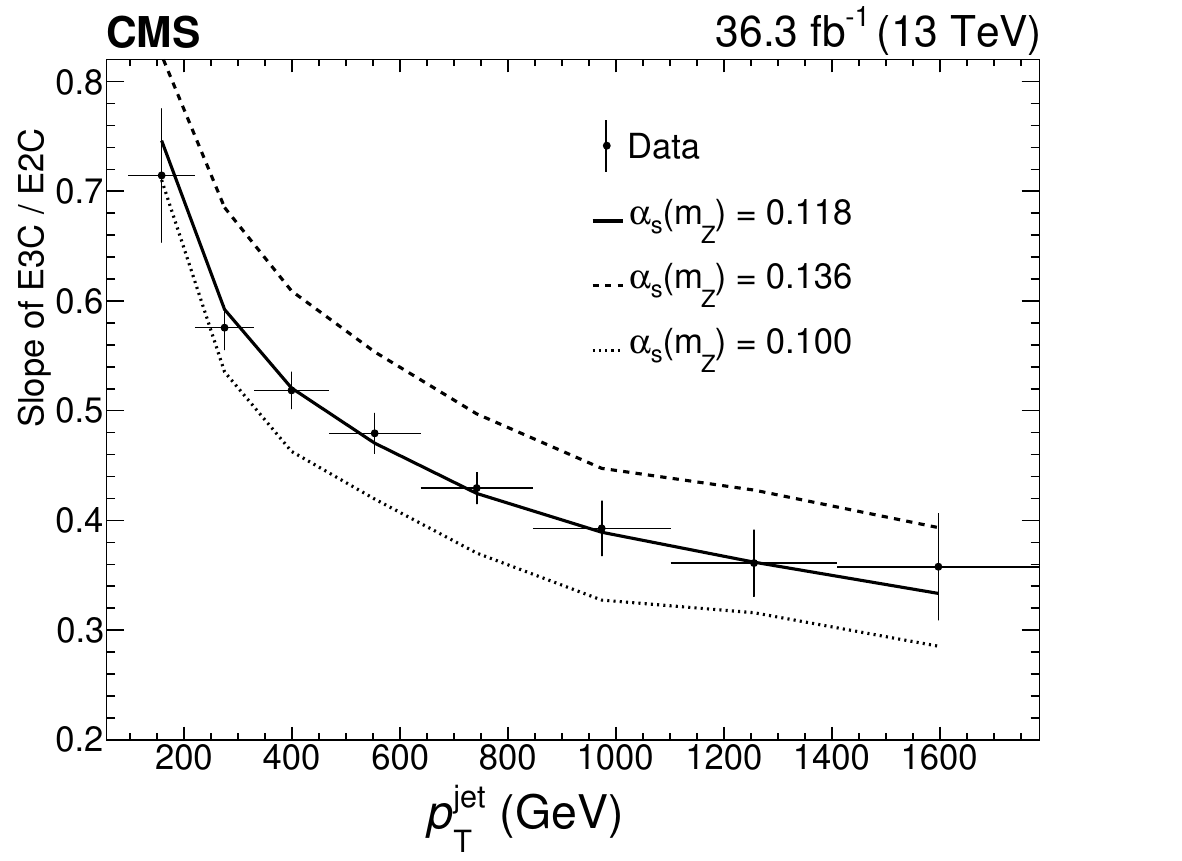}
\caption{Fitted slopes of the measured E3C/E2C ratios, in the eight jet \pt bins, 
compared to theoretical predictions for three \alpS values.}
\label{fig:slope_vs_pT}
\end{figure}

Figure~\ref{fig:dataUnfoldRatio} shows
the ratio between the E3C and E2C \xL distributions,
both measured and predicted at \NNLL\cite{Chen:2023zlx}.
The renormalization scale is set to $\ptjet R/2$ in each region, where $R=0.4$. 
This choice approximates the energy scale of the parton splitting 
and improves the convergence of the pQCD calculation~\cite{Chen:2023zlx}. 
Hadronization and UE effects are corrected using \PYTHIA{8} and \HERWIG{7} simulations,
accounting for the 1\GeV threshold on the hadron \pt.
The corrections are applied to the parton-level calculations
and are in the 5--40\% range for the E2C and E3C distributions, 
decreasing with increasing \xL and jet \pt; 
they largely cancel in the ratio, decreasing to the 0--3\% range. 
The difference between the \PYTHIA{8} and \HERWIG{7} correction factors
is considered as the nonperturbative theoretical uncertainty~\cite{CMS:2021yzl}. 
Figure~\ref{fig:slope_vs_pT} shows 
the slope of the \xL dependence of the E3C over E2C ratio,
defined as $\Delta (\mathrm{E3C/E2C}) / \Delta \log \xL$,
accounting for the covariance matrix and systematic uncertainties.
Since the slope is approximately proportional to $\alpS(Q)$~\cite{Chen:2020vvp}, 
the trend reflects the running of \alpS with jet energy.

Comparing the measured E3C/E2C ratio, as a function of \xL, with the corresponding theoretical predictions 
(using the median value of the predictions in each \xL bin), 
we obtain $\chi^2$ values as a function of \asZ. 
We consider the theoretical uncertainties described above,
except that the PS renormalization scale uncertainty is 
replaced by the \NNLL uncertainty~\cite{Chen:2023zlx}. 
Only the perturbative region is used; 
the $\xL < 0.234$ selection avoids boundary effects of the jet clustering algorithm~\cite{Lee:2022ige}. 
Since the theory normalization is unknown in the perturbative region,
a free parameter is introduced in each \pt bin,
and we only consider the shape effects of the uncertainties. 
For each \asZ value, the $\chi^2$ is defined as 
\begin{linenomath}
\ifthenelse{\boolean{cms@external}}
{ 
\begin{multline*}
\chi^2 =  [ \vec v_{\text {m}} ( \vec\theta )- \vec v_{\text{th}} ( \alpS,\vec\theta ) ]^\intercal  
V_{\text{m}}^{-1} [ \vec v_{\text {m}} (\vec\theta )- \vec v_{\text{th}} (\alpS,\vec\theta ) ] \\
+ \sum_{j}\theta_j^2 ,
\label{eqn:chi2}
\end{multline*}
}
{
\begin{equation*}
\chi^2 =  [\vec v_{\text {m}} ( \vec\theta )- \vec v_{\text{th}} ( \alpS,\vec\theta ) ]^\intercal  
V_{\text{m}} ^{-1} [ \vec v_{\text {m}} (\vec\theta )- \vec v_{\text{th}} (\alpS,\vec\theta ) ] + \sum_{j}\theta_j^2 ,
\label{eqn:chi2}
\end{equation*}
}
\end{linenomath}
where the variables $v_{\text {m}}$ and $v_{\text {th}}$ are the measured and predicted 
differential E3C/E2C ratios, respectively, 
and $V_{\text{m}}$ is the covariance matrix of the unfolded data. 
The experimental and theoretical uncertainties are considered 
through $n$ nuisance parameters, $\vec\theta=({\theta_1,\ldots}, \theta_n)$,
where $\theta_j$ is the number of standard deviations by which the 
uncertainty ``$j$" is varied.
The $\theta_j$ variation changes the shapes of the E2C and E3C distributions
simultaneously across all the bins.

The best fit value of \asZ is
$0.1229^{+0.0014}_{-0.0012}\stat {}^{+0.0030}_{-0.0033 } \thy {}^{+0.0023} _{-0.0036} (\text{exp})$, 
where theo and exp stand for theoretical and experimental systematic uncertainties, respectively. 
The central value is determined by minimizing the $\chi^2$ with respect to 
the nuisance parameters, simultaneously varied, and 
the uncertainties are given by the \asZ values that
lead to $\chi^2$ values exceeding the minimum by 1.
The high precision stems from the cancelation of most E2C and E3C systematic uncertainties 
in their ratio.
The largest sources of uncertainty are the renormalization scale 
in the theoretical calculation (2.4\%) and the energy scales of the jet constituents (2.3\%).

In summary, the two- and three-particle jet substructure observables E2C and E3C
have been measured using a sample of proton-proton collision events at $\sqrt{s} = 13\TeV$,
collected by the CMS experiment and corresponding to an integrated luminosity of 36.3\fbinv. 
A multidimensional unfolding has been performed, of the jet \pt, 
of the (largest) distance between particles in a pair or a triplet,
and of the product of their energy weights,
to compare the data with distributions simulated with 
several parton showering and hadronization models. 
This high-precision measurement of jet properties described by QCD 
can help validate future higher-order corrections in parton shower algorithms. 
The strong coupling at the \PZ boson mass
is extracted by comparing the measured E3C/E2C ratio 
with calculations at approximate next-to-next-to-leading logarithmic accuracy 
matched to a next-to-leading perturbative QCD order corrected for nonperturbative effects.
The result, $\asZ = 0.1229^{+0.0040}_{-0.0050}$, is consistent with the world average, 0.1180.
This is the most precise determination of \alpS using jet substructure techniques.
The result benefits from the development of novel jet substructure observables, 
which reduce the sensitivity to the quark-gluon composition, 
and from the availability of high-precision theoretical calculations.

\begin{acknowledgments}
We congratulate our colleagues in the CERN accelerator departments for the excellent performance of the LHC and thank the technical and administrative staffs at CERN and at other CMS institutes for their contributions to the success of the CMS effort. In addition, we gratefully acknowledge the computing centers and personnel of the Worldwide LHC Computing Grid and other centers for delivering so effectively the computing infrastructure essential to our analyses. Finally, we acknowledge the enduring support for the construction and operation of the LHC, the CMS detector, and the supporting computing infrastructure provided by the following funding agencies: SC (Armenia), BMBWF and FWF (Austria); FNRS and FWO (Belgium); CNPq, CAPES, FAPERJ, FAPERGS, and FAPESP (Brazil); MES and BNSF (Bulgaria); CERN; CAS, MoST, and NSFC (China); MINCIENCIAS (Colombia); MSES and CSF (Croatia); RIF (Cyprus); SENESCYT (Ecuador); ERC PRG, RVTT3 and MoER TK202 (Estonia); Academy of Finland, MEC, and HIP (Finland); CEA and CNRS/IN2P3 (France); SRNSF (Georgia); BMBF, DFG, and HGF (Germany); GSRI (Greece); NKFIH (Hungary); DAE and DST (India); IPM (Iran); SFI (Ireland); INFN (Italy); MSIP and NRF (Republic of Korea); MES (Latvia); LMTLT (Lithuania); MOE and UM (Malaysia); BUAP, CINVESTAV, CONACYT, LNS, SEP, and UASLP-FAI (Mexico); MOS (Montenegro); MBIE (New Zealand); PAEC (Pakistan); MES and NSC (Poland); FCT (Portugal); MESTD (Serbia); MCIN/AEI and PCTI (Spain); MOSTR (Sri Lanka); Swiss Funding Agencies (Switzerland); MST (Taipei); MHESI and NSTDA (Thailand); TUBITAK and TENMAK (Turkey); NASU (Ukraine); STFC (United Kingdom); DOE and NSF (USA).
\end{acknowledgments}

\bibliography{auto_generated}

\providecommand{\href}[2]{#2}\begingroup\raggedright\begin{thebibliography}{10}%
\makeatletter
\providecommand{\hrefCMSnoop }[0]{\@secondoftwo}%
\makeatother
\providecommand{\doi}{\texttt{doi:}\begingroup \urlstyle{tt}\Url}

\bibitem{Gross:2022hyw}
\hrefCMSnoop {}{F.~Gross { et~al.}, ``{50 years of quantum chromodynamics}'',}
  \textit{ Eur. Phys. J. C} \textbf{ 83} (2023) 1125,
  \href{http://dx.doi.org/10.1140/epjc/s10052-023-11949-2}{\doi{10.1140/epjc/s10052-023-11949-2}},
  \href{http://www.arXiv.org/abs/2212.11107}{\texttt{arXiv:2212.11107}}.

\bibitem{ATLAS:2020bbn}
\hrefCMSnoop {}{{ATLAS Collaboration}, ``Measurement of the {Lund} jet plane
  using charged particles in {13\TeV} proton-proton collisions with the {ATLAS}
  detector'',} \textit{ Phys. Rev. Lett.} \textbf{ 124} (2020) 222002,
  \href{http://dx.doi.org/10.1103/PhysRevLett.124.222002}{\doi{10.1103/PhysRevLett.124.222002}},
  \href{http://www.arXiv.org/abs/2004.03540}{\texttt{arXiv:2004.03540}}.

\bibitem{ATLAS:2017pgl}
\hrefCMSnoop {}{{ATLAS Collaboration}, ``{Measurement of jet fragmentation in
  5.02 TeV proton-lead and proton-proton collisions with the ATLAS
  detector}'',} \textit{ Nucl. Phys. A} \textbf{ 978} (2018) 65,
  \href{http://dx.doi.org/10.1016/j.nuclphysa.2018.07.006}{\doi{10.1016/j.nuclphysa.2018.07.006}},
  \href{http://www.arXiv.org/abs/1706.02859}{\texttt{arXiv:1706.02859}}.

\bibitem{ATLAS:2019mgf}
\hrefCMSnoop {}{{ATLAS Collaboration}, ``Measurement of soft-drop jet
  observables in {$\Pp\Pp$} collisions with the {ATLAS} detector at $\sqrt {s}
  = 13$ {TeV}'',} \textit{ Phys. Rev. D} \textbf{ 101} (2020) 052007,
  \href{http://dx.doi.org/10.1103/PhysRevD.101.052007}{\doi{10.1103/PhysRevD.101.052007}},
  \href{http://www.arXiv.org/abs/1912.09837}{\texttt{arXiv:1912.09837}}.

\bibitem{CMS:2021iwu}
\hrefCMSnoop {}{{CMS Collaboration}, ``Study of quark and gluon jet
  substructure in {\PZ}+jet and dijet events from pp collisions'',} \textit{
  JHEP} \textbf{ 01} (2022) 188,
  \href{http://dx.doi.org/10.1007/JHEP01(2022)188}{\doi{10.1007/JHEP01(2022)188}},
  \href{http://www.arXiv.org/abs/2109.03340}{\texttt{arXiv:2109.03340}}.

\bibitem{CMS:2017qlm}
\hrefCMSnoop {}{{CMS Collaboration}, ``Measurement of the splitting function in
  {$\Pp\Pp$} and {Pb-Pb} collisions at $\sqrt{s_{NN}}=$ 5.02 {TeV}'',} \textit{
  Phys. Rev. Lett.} \textbf{ 120} (2018) 142302,
  \href{http://dx.doi.org/10.1103/PhysRevLett.120.142302}{\doi{10.1103/PhysRevLett.120.142302}},
  \href{http://www.arXiv.org/abs/1708.09429}{\texttt{arXiv:1708.09429}}.

\bibitem{CMS:2018vzn}
\hrefCMSnoop {}{{CMS Collaboration}, ``{Measurements of the differential jet
  cross section as a function of the jet mass in dijet events from
  proton-proton collisions at $ \sqrt{s}=13 $ TeV}'',} \textit{ JHEP} \textbf{
  11} (2018) 113,
  \href{http://dx.doi.org/10.1007/JHEP11(2018)113}{\doi{10.1007/JHEP11(2018)113}},
  \href{http://www.arXiv.org/abs/1807.05974}{\texttt{arXiv:1807.05974}}.

\bibitem{ALICE:2021njq}
\hrefCMSnoop {}{{ALICE Collaboration}, ``{Measurements of the groomed and
  ungroomed jet angularities in pp collisions at $ \sqrt{s} $ = 5.02 TeV}'',}
  \textit{ JHEP} \textbf{ 05} (2022) 061,
  \href{http://dx.doi.org/10.1007/JHEP05(2022)061}{\doi{10.1007/JHEP05(2022)061}},
  \href{http://www.arXiv.org/abs/2107.11303}{\texttt{arXiv:2107.11303}}.

\bibitem{ALargeIonColliderExperiment:2021mqf}
\hrefCMSnoop {}{{ALICE Collaboration}, ``{Measurement of the groomed jet radius
  and momentum splitting fraction in pp and Pb-Pb collisions at $\sqrt{s_{NN}}
  = 5.02$ TeV}'',} \textit{ Phys. Rev. Lett.} \textbf{ 128} (2022) 102001,
  \href{http://dx.doi.org/10.1103/PhysRevLett.128.102001}{\doi{10.1103/PhysRevLett.128.102001}},
  \href{http://www.arXiv.org/abs/2107.12984}{\texttt{arXiv:2107.12984}}.

\bibitem{ALICE:2021aqk}
\hrefCMSnoop {}{{ALICE Collaboration}, ``{Direct observation of the dead-cone
  effect in quantum chromodynamics}'',} \textit{ Nature} \textbf{ 605} (2022)
  440,
  \href{http://dx.doi.org/10.1038/s41586-022-04572-w}{\doi{10.1038/s41586-022-04572-w}},
  \href{http://www.arXiv.org/abs/2106.05713}{\texttt{arXiv:2106.05713}}.
  [Erratum: \DOI{10.1038/s41586-022-05026-z}].

\bibitem{LHCb:2019qoc}
\hrefCMSnoop {}{{LHCb Collaboration}, ``{Measurement of charged hadron
  production in $\PZ$-tagged jets in proton-proton collisions at $\sqrt{s}=8$
  TeV}'',} \textit{ Phys. Rev. Lett.} \textbf{ 123} (2019) 232001,
  \href{http://dx.doi.org/10.1103/PhysRevLett.123.232001}{\doi{10.1103/PhysRevLett.123.232001}},
  \href{http://www.arXiv.org/abs/1904.08878}{\texttt{arXiv:1904.08878}}.

\bibitem{Chen:2020vvp}
\hrefCMSnoop {}{H.~Chen, I.~Moult, X.~Y. Zhang, and H.~X. Zhu, ``Rethinking
  jets with energy correlators: tracks, resummation, and analytic
  continuation'',} \textit{ Phys. Rev. D} \textbf{ 102} (2020) 054012,
  \href{http://dx.doi.org/10.1103/PhysRevD.102.054012}{\doi{10.1103/PhysRevD.102.054012}},
  \href{http://www.arXiv.org/abs/2004.11381}{\texttt{arXiv:2004.11381}}.

\bibitem{Basham:1978bw}
\hrefCMSnoop {}{C.~L. Basham, L.~S. Brown, S.~D. Ellis, and S.~T. Love,
  ``Energy correlations in electron - positron annihilation: Testing {QCD}'',}
  \textit{ Phys. Rev. Lett.} \textbf{ 41} (1978) 1585,
  \href{http://dx.doi.org/10.1103/PhysRevLett.41.1585}{\doi{10.1103/PhysRevLett.41.1585}}.

\bibitem{PLUTO:1981gcc}
\hrefCMSnoop {}{{PLUTO} Collaboration, ``Energy-energy correlations in
  {\EE}annihilation into hadrons'',} \textit{ Phys. Lett. B} \textbf{ 99}
  (1981) 292,
  \href{http://dx.doi.org/10.1016/0370-2693(81)91128-X}{\doi{10.1016/0370-2693(81)91128-X}}.

\bibitem{L3:1992btq}
\hrefCMSnoop {}{{L3} Collaboration, ``{Determination of $\alpha_S$ from
  hadronic event shapes measured on the $\PZ^0$ resonance}'',} \textit{ Phys.
  Lett. B} \textbf{ 284} (1992) 471,
  \href{http://dx.doi.org/10.1016/0370-2693(92)90463-E}{\doi{10.1016/0370-2693(92)90463-E}}.

\bibitem{OPAL:1991uui}
\hrefCMSnoop {}{{OPAL} Collaboration, ``{An improved measurement of
  $\alpha_S(m_{\PZ^0})$ using energy correlations with the OPAL detector at
  LEP}'',} \textit{ Phys. Lett. B} \textbf{ 276} (1992) 547,
  \href{http://dx.doi.org/10.1016/0370-2693(92)91681-X}{\doi{10.1016/0370-2693(92)91681-X}}.

\bibitem{DELPHI:1992qrr}
\hrefCMSnoop {}{{DELPHI} Collaboration, ``{Determination of $\alpha_S$ in
  second order QCD from hadronic \PZ decays}'',} \textit{ Z. Phys. C} \textbf{
  54} (1992) 55,
  \href{http://dx.doi.org/10.1007/BF01881708}{\doi{10.1007/BF01881708}}.

\bibitem{Hofman:2008ar}
\hrefCMSnoop {}{D.~M. Hofman and J.~Maldacena, ``Conformal collider physics:
  Energy and charge correlations'',} \textit{ JHEP} \textbf{ 05} (2008) 012,
  \href{http://dx.doi.org/10.1088/1126-6708/2008/05/012}{\doi{10.1088/1126-6708/2008/05/012}},
  \href{http://www.arXiv.org/abs/0803.1467}{\texttt{arXiv:0803.1467}}.

\bibitem{Salam:2010nqg}
\hrefCMSnoop {}{G.~P. Salam, ``{Towards Jetography}'',} \textit{ Eur. Phys. J.
  C} \textbf{ 67} (2010) 637,
  \href{http://dx.doi.org/10.1140/epjc/s10052-010-1314-6}{\doi{10.1140/epjc/s10052-010-1314-6}},
  \href{http://www.arXiv.org/abs/0906.1833}{\texttt{arXiv:0906.1833}}.

\bibitem{Sveshnikov:1995vi}
\hrefCMSnoop {}{N.~A. Sveshnikov and F.~V. Tkachov, ``{Jets and quantum field
  theory}'',} \textit{ Phys. Lett. B} \textbf{ 382} (1996) 403,
  \href{http://dx.doi.org/10.1016/0370-2693(96)00558-8}{\doi{10.1016/0370-2693(96)00558-8}},
  \href{http://www.arXiv.org/abs/hep-ph/9512370}{\texttt{arXiv:hep-ph/9512370}}.

\bibitem{Belitsky:2013xxa}
A.~V. Belitsky\hrefCMSnoop {}{ { et~al.}, ``{From correlation functions to
  event shapes}'',} \textit{ Nucl. Phys. B} \textbf{ 884} (2014) 305,
  \href{http://dx.doi.org/10.1016/j.nuclphysb.2014.04.020}{\doi{10.1016/j.nuclphysb.2014.04.020}},
  \href{http://www.arXiv.org/abs/1309.0769}{\texttt{arXiv:1309.0769}}.

\bibitem{Belitsky:2013bja}
A.~V. Belitsky\hrefCMSnoop {}{ { et~al.}, ``{Event shapes in $\mathcal{N} = 4$
  super-Yang-Mills theory}'',} \textit{ Nucl. Phys. B} \textbf{ 884} (2014)
  206,
  \href{http://dx.doi.org/10.1016/j.nuclphysb.2014.04.019}{\doi{10.1016/j.nuclphysb.2014.04.019}},
  \href{http://www.arXiv.org/abs/1309.1424}{\texttt{arXiv:1309.1424}}.

\bibitem{Kologlu:2019mfz}
\hrefCMSnoop {}{M.~Kologlu, P.~Kravchuk, D.~Simmons-Duffin, and A.~Zhiboedov,
  ``{The light-ray OPE and conformal colliders}'',} \textit{ JHEP} \textbf{ 01}
  (2021) 128,
  \href{http://dx.doi.org/10.1007/JHEP01(2021)128}{\doi{10.1007/JHEP01(2021)128}},
  \href{http://www.arXiv.org/abs/1905.01311}{\texttt{arXiv:1905.01311}}.

\bibitem{Dixon:2019uzg}
\hrefCMSnoop {}{L.~J. Dixon, I.~Moult, and H.~X. Zhu, ``{Collinear limit of the
  energy-energy correlator}'',} \textit{ Phys. Rev. D} \textbf{ 100} (2019)
  014009,
  \href{http://dx.doi.org/10.1103/PhysRevD.100.014009}{\doi{10.1103/PhysRevD.100.014009}},
  \href{http://www.arXiv.org/abs/1905.01310}{\texttt{arXiv:1905.01310}}.

\bibitem{Lee:2022ige}
\hrefCMSnoop {}{K.~Lee, B.~Me{\c{c}}aj, and I.~Moult, ``Conformal colliders
  meet the {LHC}'',} 2022.
  \href{http://www.arXiv.org/abs/2205.03414}{\texttt{arXiv:2205.03414}}.

\bibitem{Chen:2023zlx}
W.~Chen\hrefCMSnoop {}{ { et~al.}, ``{NNLL resummation for projected
  three-point energy correlator}'',} \textit{ JHEP} \textbf{ 05} (2024) 043,
  \href{http://dx.doi.org/10.1007/JHEP05(2024)043}{\doi{10.1007/JHEP05(2024)043}},
  \href{http://www.arXiv.org/abs/2307.07510}{\texttt{arXiv:2307.07510}}.

\bibitem{Proceedings:2018jsb}
\hrefCMSnoop {}{J.~R. Andersen { et~al.}, ``{Les Houches} 2017: Physics at
  {TeV} colliders standard model working group report'',} in \textit{ Proc.
  Standard Model Working Group of the 2017 Les Houches Workshop}.
\newblock 2018.
\newblock
  \href{http://www.arXiv.org/abs/1803.07977}{\texttt{arXiv:1803.07977}}.

\bibitem{Salam:2017qdl}
\hrefCMSnoop {}{G.~P. Salam, ``{The strong coupling: a theoretical
  perspective}'',} CERN theory report CERN-TH-2017-268, 2017.
\newblock
  \href{http://dx.doi.org/10.1142/9789813238053_0007}{\doi{10.1142/9789813238053_0007}},
  \href{http://www.arXiv.org/abs/1712.05165}{\texttt{arXiv:1712.05165}}.

\bibitem{Moch:2014tta}
\hrefCMSnoop {}{S.~Moch { et~al.}, ``{High precision fundamental constants at
  the TeV scale}'',} 2014.
  \href{http://www.arXiv.org/abs/1405.4781}{\texttt{arXiv:1405.4781}}.

\bibitem{LeBlanc:2022bwd}
\hrefCMSnoop {}{M.~LeBlanc, B.~Nachman, and C.~Sauer, ``{Going off topics to
  demix quark and gluon jets in $\alpha_{S}$ extractions}'',} \textit{ JHEP}
  \textbf{ 02} (2023) 150,
  \href{http://dx.doi.org/10.1007/JHEP02(2023)150}{\doi{10.1007/JHEP02(2023)150}},
  \href{http://www.arXiv.org/abs/2206.10642}{\texttt{arXiv:2206.10642}}.

\bibitem{ParticleDataGroup:2022pth}
\hrefCMSnoop {}{{Particle Data Group}, R.~L. Workman { et~al.}, ``Review of
  particle physics'',} \textit{ Prog. Theor. Exp. Phys.} \textbf{ 2022} (2022)
  083C01,
  \href{http://dx.doi.org/10.1093/ptep/ptac097}{\doi{10.1093/ptep/ptac097}}.
  and 2023 update.

\bibitem{ATLAS:2023lhg}
\hrefCMSnoop {}{{ATLAS Collaboration}, ``{A precise determination of the
  strong-coupling constant from the recoil of $\PZ$ bosons with the ATLAS
  experiment at $\sqrt{s} = 8$ TeV}'',} 2023.
  \href{http://www.arXiv.org/abs/2309.12986}{\texttt{arXiv:2309.12986}}.
Submitted to \textit{Nature Phys.}

\bibitem{CMS:2021yzl}
\hrefCMSnoop {}{{CMS Collaboration}, ``{Measurement and QCD analysis of
  double-differential inclusive jet cross sections in proton-proton collisions
  at $ \sqrt{s} $ = 13 TeV}'',} \textit{ JHEP} \textbf{ 02} (2022) 142,
  \href{http://dx.doi.org/10.1007/JHEP02(2022)142}{\doi{10.1007/JHEP02(2022)142}},
  \href{http://www.arXiv.org/abs/2111.10431}{\texttt{arXiv:2111.10431}}.
  [Addendum: \DOI{10.1007/JHEP12(2022)035}].

\bibitem{ATLAS:2023tgo}
\hrefCMSnoop {}{{ATLAS Collaboration}, ``{Determination of the strong coupling
  constant from transverse energy-energy correlations in multijet events at
  $\sqrt{s} = 13$\TeV with the ATLAS detector}'',} \textit{ JHEP} \textbf{ 07}
  (2023) 085,
  \href{http://dx.doi.org/10.1007/JHEP07(2023)085}{\doi{10.1007/JHEP07(2023)085}},
  \href{http://www.arXiv.org/abs/2301.09351}{\texttt{arXiv:2301.09351}}.

\bibitem{CMS:2018fks}
\hrefCMSnoop {}{{CMS Collaboration}, ``{Measurement of the
  $\mathrm{t}\overline{\mathrm{t}}$ production cross section, the top quark
  mass, and the strong coupling constant using dilepton events in pp collisions
  at $\sqrt{s} = 13$\TeV}'',} \textit{ Eur. Phys. J. C} \textbf{ 79} (2019)
  368,
  \href{http://dx.doi.org/10.1140/epjc/s10052-019-6863-8}{\doi{10.1140/epjc/s10052-019-6863-8}},
  \href{http://www.arXiv.org/abs/1812.10505}{\texttt{arXiv:1812.10505}}.

\bibitem{dEnterria:2019aat}
\hrefCMSnoop {}{D.~d'Enterria and A.~Poldaru, ``{Extraction of the strong
  coupling $\asZ$ from a combined NNLO analysis of inclusive electroweak boson
  cross sections at hadron colliders}'',} \textit{ JHEP} \textbf{ 06} (2020)
  016,
  \href{http://dx.doi.org/10.1007/JHEP06(2020)016}{\doi{10.1007/JHEP06(2020)016}},
  \href{http://www.arXiv.org/abs/1912.11733}{\texttt{arXiv:1912.11733}}.

\bibitem{Klijnsma:2017eqp}
\hrefCMSnoop {}{T.~Klijnsma, S.~Bethke, G.~Dissertori, and G.~P. Salam,
  ``{Determination of the strong coupling constant $\asZ$ from measurements of
  the total cross section for top-antitop quark production}'',} \textit{ Eur.
  Phys. J. C} \textbf{ 77} (2017) 778,
  \href{http://dx.doi.org/10.1140/epjc/s10052-017-5340-5}{\doi{10.1140/epjc/s10052-017-5340-5}},
  \href{http://www.arXiv.org/abs/1708.07495}{\texttt{arXiv:1708.07495}}.

\bibitem{Abbate:2010xh}
R.~Abbate\hrefCMSnoop {}{ { et~al.}, ``{Thrust at $\text{N}^{3}\text{LL}$ with
  power corrections and a precision global fit for $\asZ$}'',} \textit{ Phys.
  Rev. D} \textbf{ 83} (2011) 074021,
  \href{http://dx.doi.org/10.1103/PhysRevD.83.074021}{\doi{10.1103/PhysRevD.83.074021}},
  \href{http://www.arXiv.org/abs/1006.3080}{\texttt{arXiv:1006.3080}}.

\bibitem{Gehrmann:2012sc}
\hrefCMSnoop {}{T.~Gehrmann, G.~Luisoni, and P.~F. Monni, ``{Power corrections
  in the dispersive model for a determination of the strong coupling constant
  from the thrust distribution}'',} \textit{ Eur. Phys. J. C} \textbf{ 73}
  (2013) 2265,
  \href{http://dx.doi.org/10.1140/epjc/s10052-012-2265-x}{\doi{10.1140/epjc/s10052-012-2265-x}},
  \href{http://www.arXiv.org/abs/1210.6945}{\texttt{arXiv:1210.6945}}.

\bibitem{Hoang:2015hka}
\hrefCMSnoop {}{A.~H. Hoang, D.~W. Kolodrubetz, V.~Mateu, and I.~W. Stewart,
  ``{Precise determination of $\alpS$ from the $C$-parameter distribution}'',}
  \textit{ Phys. Rev. D} \textbf{ 91} (2015) 094018,
  \href{http://dx.doi.org/10.1103/PhysRevD.91.094018}{\doi{10.1103/PhysRevD.91.094018}},
  \href{http://www.arXiv.org/abs/1501.04111}{\texttt{arXiv:1501.04111}}.

\bibitem{CMS:2018ypj}
\hrefCMSnoop {}{{CMS Collaboration}, ``{Measurement of jet substructure
  observables in $\mathrm{t\overline{t}}$ events from proton-proton collisions
  at $\sqrt{s}=$ 13 TeV}'',} \textit{ Phys. Rev. D} \textbf{ 98} (2018) 092014,
  \href{http://dx.doi.org/10.1103/PhysRevD.98.092014}{\doi{10.1103/PhysRevD.98.092014}},
  \href{http://www.arXiv.org/abs/1808.07340}{\texttt{arXiv:1808.07340}}.

\bibitem{Gribov:1972ri}
\hrefCMSnoop {}{V.~N. Gribov and L.~N. Lipatov, ``{Deep inelastic ep scattering
  in perturbation theory}'',} \textit{ Sov. J. Nucl. Phys.} \textbf{ 15} (1972)
  438.

\bibitem{Dokshitzer:1977sg}
\hrefCMSnoop {}{Y.~L. Dokshitzer, ``Calculation of the structure functions for
  deep inelastic scattering and {$\EE$} annihilation by perturbation theory in
  quantum chromodynamics.'',} \textit{ Sov. Phys. JETP} \textbf{ 46} (1977)
  641.

\bibitem{Marchesini:1994wr}
\hrefCMSnoop {}{G.~Marchesini, ``{QCD coherence in the structure function and
  associated distributions at small x}'',} \textit{ Nucl. Phys. B} \textbf{
  445} (1995) 49,
  \href{http://dx.doi.org/10.1016/0550-3213(95)00149-M}{\doi{10.1016/0550-3213(95)00149-M}},
  \href{http://www.arXiv.org/abs/hep-ph/9412327}{\texttt{arXiv:hep-ph/9412327}}.

\bibitem{Altarelli:1977zs}
\hrefCMSnoop {}{G.~Altarelli and G.~Parisi, ``Asymptotic freedom in parton
  language'',} \textit{ Nucl. Phys. B} \textbf{ 126} (1977) 298,
  \href{http://dx.doi.org/10.1016/0550-3213(77)90384-4}{\doi{10.1016/0550-3213(77)90384-4}}.

\bibitem{CMS-LUM-17-003}
\hrefCMSnoop {}{{CMS Collaboration}, ``Precision luminosity measurement in
  proton-proton collisions at $\sqrt{s} =$ 13 {TeV} in 2015 and 2016 at
  {CMS}'',} \textit{ Eur. Phys. J. C} \textbf{ 81} (2021) 800,
  \href{http://dx.doi.org/10.1140/epjc/s10052-021-09538-2}{\doi{10.1140/epjc/s10052-021-09538-2}},
  \href{http://www.arXiv.org/abs/2104.01927}{\texttt{arXiv:2104.01927}}.

\bibitem{CMS:2008xjf}
\hrefCMSnoop {}{{CMS Collaboration}, ``The {CMS} experiment at the {CERN}
  {LHC}'',} \textit{ JINST} \textbf{ 3} (2008) S08004,
  \href{http://dx.doi.org/10.1088/1748-0221/3/08/S08004}{\doi{10.1088/1748-0221/3/08/S08004}}.

\bibitem{CMS:2020cmk}
\hrefCMSnoop {}{{CMS Collaboration}, ``{Performance of the CMS Level-1 trigger
  in proton-proton collisions at $\sqrt{s} = 13$\,TeV}'',} \textit{ JINST}
  \textbf{ 15} (2020) P10017,
  \href{http://dx.doi.org/10.1088/1748-0221/15/10/P10017}{\doi{10.1088/1748-0221/15/10/P10017}},
  \href{http://www.arXiv.org/abs/2006.10165}{\texttt{arXiv:2006.10165}}.

\bibitem{CMS:2016ngn}
\hrefCMSnoop {}{{CMS Collaboration}, ``{The CMS trigger system}'',} \textit{
  JINST} \textbf{ 12} (2017) P01020,
  \href{http://dx.doi.org/10.1088/1748-0221/12/01/P01020}{\doi{10.1088/1748-0221/12/01/P01020}},
\href{http://www.arXiv.org/abs/1609.02366}{\texttt{arXiv:1609.02366}}.

\bibitem{CMS:2020uim}
\hrefCMSnoop {}{{CMS Collaboration}, ``Electron and photon reconstruction and
  identification with the {CMS} experiment at the {CERN} {LHC}'',} \textit{
  JINST} \textbf{ 16} (2021) P05014,
  \href{http://dx.doi.org/10.1088/1748-0221/16/05/P05014}{\doi{10.1088/1748-0221/16/05/P05014}},
  \href{http://www.arXiv.org/abs/2012.06888}{\texttt{arXiv:2012.06888}}.

\bibitem{CMS:2018rym}
\hrefCMSnoop {}{{CMS Collaboration}, ``{Performance of the CMS muon detector
  and muon reconstruction with proton-proton collisions at $\sqrt{s}=$ 13
  TeV}'',} \textit{ JINST} \textbf{ 13} (2018) P06015,
  \href{http://dx.doi.org/10.1088/1748-0221/13/06/P06015}{\doi{10.1088/1748-0221/13/06/P06015}},
  \href{http://www.arXiv.org/abs/1804.04528}{\texttt{arXiv:1804.04528}}.

\bibitem{CMS:2014pgm}
\hrefCMSnoop {}{{CMS Collaboration}, ``{Description and performance of track
  and primary-vertex reconstruction with the CMS tracker}'',} \textit{ JINST}
  \textbf{ 9} (2014) P10009,
  \href{http://dx.doi.org/10.1088/1748-0221/9/10/P10009}{\doi{10.1088/1748-0221/9/10/P10009}},
  \href{http://www.arXiv.org/abs/1405.6569}{\texttt{arXiv:1405.6569}}.

\bibitem{CMS:2019ctu}
\hrefCMSnoop {}{{CMS Collaboration}, ``Performance of missing transverse
  momentum reconstruction in proton-proton collisions at $\sqrt{s} = 13$\,{TeV}
  using the {CMS} detector'',} \textit{ JINST} \textbf{ 14} (2019) P07004,
  \href{http://dx.doi.org/10.1088/1748-0221/14/07/P07004}{\doi{10.1088/1748-0221/14/07/P07004}},
\href{http://www.arXiv.org/abs/1903.06078}{\texttt{arXiv:1903.06078}}.

\bibitem{CMS:2017yfk}
\hrefCMSnoop {}{{CMS Collaboration}, ``Particle-flow reconstruction and global
  event description with the {CMS} detector'',} \textit{ JINST} \textbf{ 12}
  (2017) P10003,
  \href{http://dx.doi.org/10.1088/1748-0221/12/10/P10003}{\doi{10.1088/1748-0221/12/10/P10003}},
\href{http://www.arXiv.org/abs/1706.04965}{\texttt{arXiv:1706.04965}}.

\bibitem{CMS:2018jrd}
\hrefCMSnoop {}{{CMS Collaboration}, ``{Performance of reconstruction and
  identification of $\tau$ leptons decaying to hadrons and $\nu_\tau$ in pp
  collisions at $\sqrt{s}=$ 13 TeV}'',} \textit{ JINST} \textbf{ 13} (2018)
  P10005,
  \href{http://dx.doi.org/10.1088/1748-0221/13/10/P10005}{\doi{10.1088/1748-0221/13/10/P10005}},
  \href{http://www.arXiv.org/abs/1809.02816}{\texttt{arXiv:1809.02816}}.

\bibitem{CMS:2016lmd}
\hrefCMSnoop {}{{CMS Collaboration}, ``{Jet energy scale and resolution in the
  CMS experiment in pp collisions at 8 TeV}'',} \textit{ JINST} \textbf{ 12}
  (2017) P02014,
  \href{http://dx.doi.org/10.1088/1748-0221/12/02/P02014}{\doi{10.1088/1748-0221/12/02/P02014}},
\href{http://www.arXiv.org/abs/1607.03663}{\texttt{arXiv:1607.03663}}.

\bibitem{Cacciari:2008gp}
\hrefCMSnoop {}{M.~Cacciari, G.~P. Salam, and G.~Soyez, ``The anti-\kt jet
  clustering algorithm'',} \textit{ JHEP} \textbf{ 04} (2008) 063,
  \href{http://dx.doi.org/10.1088/1126-6708/2008/04/063}{\doi{10.1088/1126-6708/2008/04/063}},
  \href{http://www.arXiv.org/abs/0802.1189}{\texttt{arXiv:0802.1189}}.

\bibitem{Cacciari:2011ma}
\hrefCMSnoop {}{M.~Cacciari, G.~P. Salam, and G.~Soyez, ``{FastJet user
  manual}'',} \textit{ Eur. Phys. J. C} \textbf{ 72} (2012) 1896,
  \href{http://dx.doi.org/10.1140/epjc/s10052-012-1896-2}{\doi{10.1140/epjc/s10052-012-1896-2}},
\href{http://www.arXiv.org/abs/1111.6097}{\texttt{arXiv:1111.6097}}.

\bibitem{DAgostini}
\hrefCMSnoop {}{G.~D'Agostini, ``{A multidimensional unfolding method based on
  Bayes' theorem}'',} \textit{ Nucl. Instrum. Meth. A} \textbf{ 362} (1995)
  487,
  \href{http://dx.doi.org/10.1016/0168-9002(95)00274-X}{\doi{10.1016/0168-9002(95)00274-X}}.

\bibitem{Adye:2011gm}
\hrefCMSnoop {}{T.~Adye, ``{Unfolding algorithms and tests using RooUnfold}'',}
  in \textit{ {PHYSTAT 2011}}, p.~313.
\newblock CERN, Geneva, 2011.
\newblock \href{http://www.arXiv.org/abs/1105.1160}{\texttt{arXiv:1105.1160}}.
\newblock
  \href{http://dx.doi.org/10.5170/CERN-2011-006.313}{\doi{10.5170/CERN-2011-006.313}}.

\bibitem{Brenner:2019lmf}
L.~Brenner\hrefCMSnoop {}{ { et~al.}, ``{Comparison of unfolding methods using
  RooFitUnfold}'',} \textit{ Int. J. Mod. Phys. A} \textbf{ 35} (2020) 2050145,
  \href{http://dx.doi.org/10.1142/S0217751X20501456}{\doi{10.1142/S0217751X20501456}},
  \href{http://www.arXiv.org/abs/1910.14654}{\texttt{arXiv:1910.14654}}.

\bibitem{Sjostrand:2014zea}
T.~Sj{\"o}strand\hrefCMSnoop {}{ { et~al.}, ``An introduction to {\PYTHIA}
  8.2'',} \textit{ Comput. Phys. Commun.} \textbf{ 191} (2015) 159,
  \href{http://dx.doi.org/10.1016/j.cpc.2015.01.024}{\doi{10.1016/j.cpc.2015.01.024}},
\href{http://www.arXiv.org/abs/1410.3012}{\texttt{arXiv:1410.3012}}.

\bibitem{Bahr:2008pv}
\hrefCMSnoop {}{M.~Bahr { et~al.}, ``{\HERWIG}++ physics and manual'',}
  \textit{ Eur. Phys. J. C} \textbf{ 58} (2008) 639,
  \href{http://dx.doi.org/10.1140/epjc/s10052-008-0798-9}{\doi{10.1140/epjc/s10052-008-0798-9}},
  \href{http://www.arXiv.org/abs/0803.0883}{\texttt{arXiv:0803.0883}}.

\bibitem{Bellm:2015jjp}
\hrefCMSnoop {}{J.~Bellm { et~al.}, ``{\HERWIG 7.0/\HERWIG++ 3.0 release
  note}'',} \textit{ Eur. Phys. J. C} \textbf{ 76} (2016) 196,
  \href{http://dx.doi.org/10.1140/epjc/s10052-016-4018-8}{\doi{10.1140/epjc/s10052-016-4018-8}},
  \href{http://www.arXiv.org/abs/1512.01178}{\texttt{arXiv:1512.01178}}.

\bibitem{Alwall:2014hca}
J.~Alwall\hrefCMSnoop {}{ { et~al.}, ``{The automated computation of tree-level
  and next-to-leading order differential cross sections, and their matching to
  parton shower simulations}'',} \textit{ JHEP} \textbf{ 07} (2014) 079,
  \href{http://dx.doi.org/10.1007/JHEP07(2014)079}{\doi{10.1007/JHEP07(2014)079}},
  \href{http://www.arXiv.org/abs/1405.0301}{\texttt{arXiv:1405.0301}}.

\bibitem{Alwall:2007fs}
\hrefCMSnoop {}{J.~Alwall { et~al.}, ``{Comparative study of various algorithms
  for the merging of parton showers and matrix elements in hadronic
  collisions}'',} \textit{ Eur. Phys. J. C} \textbf{ 53} (2008) 473,
  \href{http://dx.doi.org/10.1140/epjc/s10052-007-0490-5}{\doi{10.1140/epjc/s10052-007-0490-5}},
  \href{http://www.arXiv.org/abs/0706.2569}{\texttt{arXiv:0706.2569}}.

\bibitem{Ball:2014uwa}
\hrefCMSnoop {}{{NNPDF} Collaboration, ``Parton distributions for the {LHC Run
  II}'',} \textit{ JHEP} \textbf{ 04} (2015) 040,
  \href{http://dx.doi.org/10.1007/JHEP04(2015)040}{\doi{10.1007/JHEP04(2015)040}},
\href{http://www.arXiv.org/abs/1410.8849}{\texttt{arXiv:1410.8849}}.

\bibitem{Agostinelli:2002hh}
\hrefCMSnoop {}{{GEANT4} Collaboration, ``{\GEANTfour}---a simulation
  toolkit'',} \textit{ Nucl. Instrum. Meth. A} \textbf{ 506} (2003) 250,
\href{http://dx.doi.org/10.1016/S0168-9002(03)01368-8}{\doi{10.1016/S0168-9002(03)01368-8}}.

\bibitem{CMS:2020ebo}
\hrefCMSnoop {}{{CMS Collaboration}, ``{Pileup mitigation at CMS in 13\TeV
  data}'',} \textit{ JINST} \textbf{ 15} (2020) P09018,
  \href{http://dx.doi.org/10.1088/1748-0221/15/09/P09018}{\doi{10.1088/1748-0221/15/09/P09018}},
  \href{http://www.arXiv.org/abs/2003.00503}{\texttt{arXiv:2003.00503}}.

\bibitem{suppMatBib2}
\hrefCMSnoop {}{``Placeholder for supplemental material
  ({Appendix~\ref{app:suppMat}} in preprint version): Additional analysis
  figures and referenced material
  \cite{Fischer:2016vfv,Hoche:2015sya,Sherpa:2019gpd}'',} 2024.
\newblock [URL will be provided by publisher at publication].

\bibitem{hepdata}
\hrefCMSnoop {}{}{HEPD}ata record for this analysis, 2024.
\newblock
  \href{http://dx.doi.org/10.17182/hepdata.147275}{\doi{10.17182/hepdata.147275}}.

\bibitem{Mrenna:2016sih}
\hrefCMSnoop {}{S.~Mrenna and P.~Skands, ``{Automated parton-shower variations
  in {$\PYTHIA8$}}'',} \textit{ Phys. Rev. D} \textbf{ 94} (2016) 074005,
  \href{http://dx.doi.org/10.1103/PhysRevD.94.074005}{\doi{10.1103/PhysRevD.94.074005}},
  \href{http://www.arXiv.org/abs/1605.08352}{\texttt{arXiv:1605.08352}}.

\bibitem{Skands:2014pea}
\hrefCMSnoop {}{P.~Skands, S.~Carrazza, and J.~Rojo, ``{Tuning \PYTHIA 8.1: the
  Monash 2013 tune}'',} \textit{ Eur. Phys. J. C} \textbf{ 74} (2014) 3024,
  \href{http://dx.doi.org/10.1140/epjc/s10052-014-3024-y}{\doi{10.1140/epjc/s10052-014-3024-y}},
  \href{http://www.arXiv.org/abs/1404.5630}{\texttt{arXiv:1404.5630}}.

\bibitem{Komiske:2022enw}
\hrefCMSnoop {}{P.~T. Komiske, I.~Moult, J.~Thaler, and H.~X. Zhu, ``Analyzing
  {$N$}-point energy correlators inside jets with {CMS} open data'',} \textit{
  Phys. Rev. Lett.} \textbf{ 130} (2023) 051901,
  \href{http://dx.doi.org/10.1103/PhysRevLett.130.051901}{\doi{10.1103/PhysRevLett.130.051901}},
  \href{http://www.arXiv.org/abs/2201.07800}{\texttt{arXiv:2201.07800}}.

\bibitem{Fischer:2016vfv}
\hrefCMSnoop {}{N.~Fischer, S.~Prestel, M.~Ritzmann, and P.~Skands, ``{\VINCIA}
  for hadron colliders'',} \textit{ Eur. Phys. J. C} \textbf{ 76} (2016) 589,
  \href{http://dx.doi.org/10.1140/epjc/s10052-016-4429-6}{\doi{10.1140/epjc/s10052-016-4429-6}},
  \href{http://www.arXiv.org/abs/1605.06142}{\texttt{arXiv:1605.06142}}.

\bibitem{Hoche:2015sya}
\hrefCMSnoop {}{S.~H{\"o}che and S.~Prestel, ``{The midpoint between dipole and
  parton showers}'',} \textit{ Eur. Phys. J. C} \textbf{ 75} (2015) 461,
  \href{http://dx.doi.org/10.1140/epjc/s10052-015-3684-2}{\doi{10.1140/epjc/s10052-015-3684-2}},
  \href{http://www.arXiv.org/abs/1506.05057}{\texttt{arXiv:1506.05057}}.

\bibitem{Sherpa:2019gpd}
\hrefCMSnoop {}{{Sherpa} Collaboration, ``Event generation with {\SHERPA}
  2.2'',} \textit{ SciPost Phys.} \textbf{ 7} (2019) 034,
  \href{http://dx.doi.org/10.21468/SciPostPhys.7.3.034}{\doi{10.21468/SciPostPhys.7.3.034}},
  \href{http://www.arXiv.org/abs/1905.09127}{\texttt{arXiv:1905.09127}}.

\end{thebibliography}\endgroup
\ifthenelse{\boolean{cms@external}}{\nocite{Fischer:2016vfv,Hoche:2015sya,Sherpa:2019gpd} }{
\clearpage
\appendix
\numberwithin{table}{section}
\numberwithin{figure}{section}
\section{Additional analysis figures}
\label{app:suppMat}
Figures~\ref{fig:E2C_sup} and~\ref{fig:E3C_sup} show the measured (unfolded) E2C and E3C distributions, compared to various MC predictions in the eight \pt regions of the analysis. The MC predictions include 
\PYTHIA{8.240} using \VINCIA\cite{Fischer:2016vfv} and \DIRE\cite{Hoche:2015sya} showers, with the CP5 tune;
\HERWIG{7.1.4} using dipole showers, with the CH3 tune; 
and \SHERPA{2.2}~\cite{Sherpa:2019gpd}, with its default tune.
The value of \asZ is set to 0.118, with its energy evolution computed at two-loop order. 
The E3C/E2C ratios are compared with MC predictions in Fig.~\ref{fig:ratio_sup} and with theoretical predictions in Fig.~\ref{fig:ratio_theo_sup}.
The $\chi^2$ values determined from the fit of measured data as a function of \asZ are shown in Fig.~\ref{fig:chi2}.
The statistical correlation matrices of the unfolded E2C and E3C distributions, and of their ratio, are shown in Figs.~\ref{fig:E2C_cor}--\ref{fig:ratio_cor}.

\begin{figure*}[htbp]
\centering
\includegraphics[width=\linewidth]{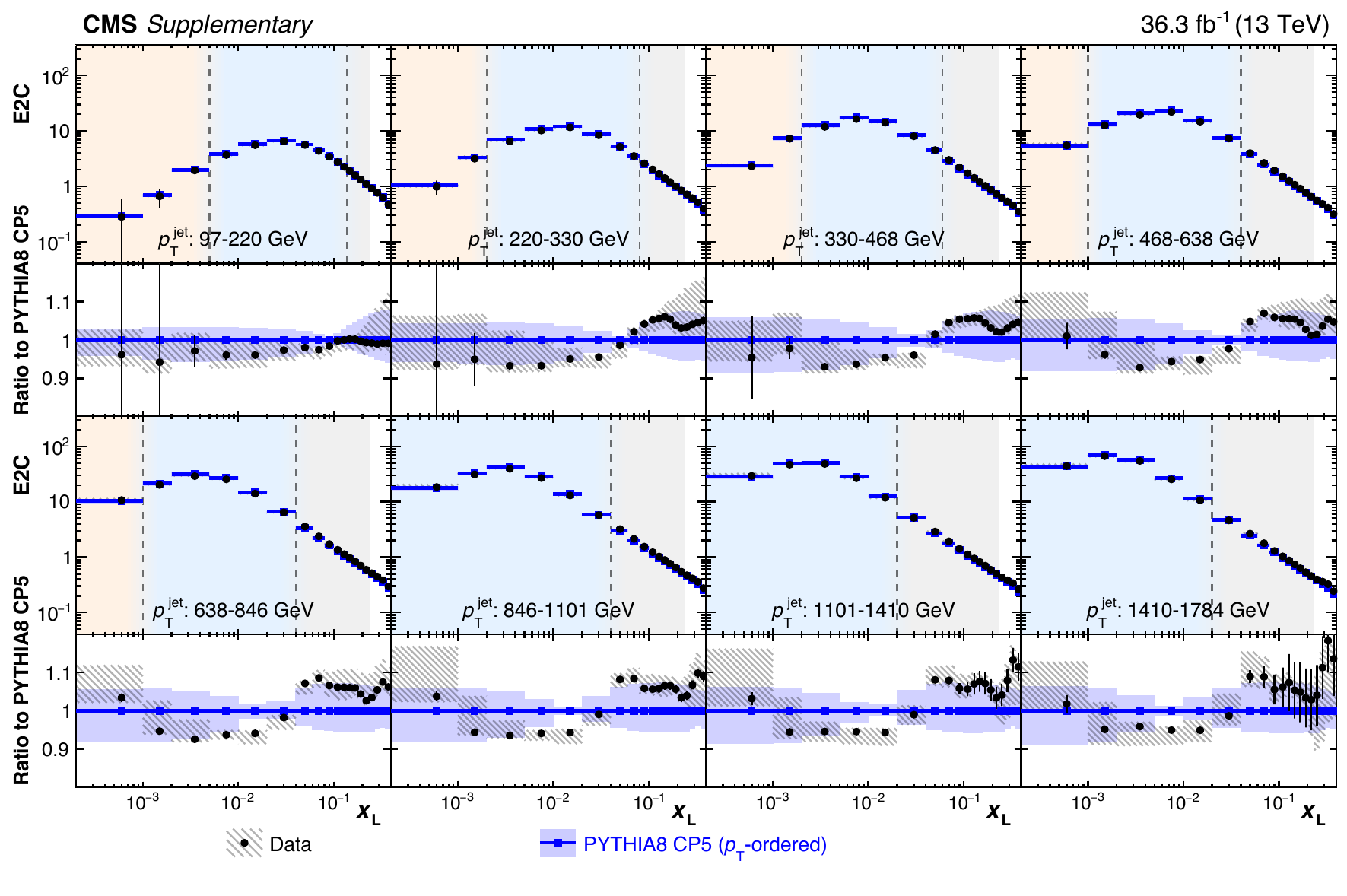}
\includegraphics[width=\linewidth]{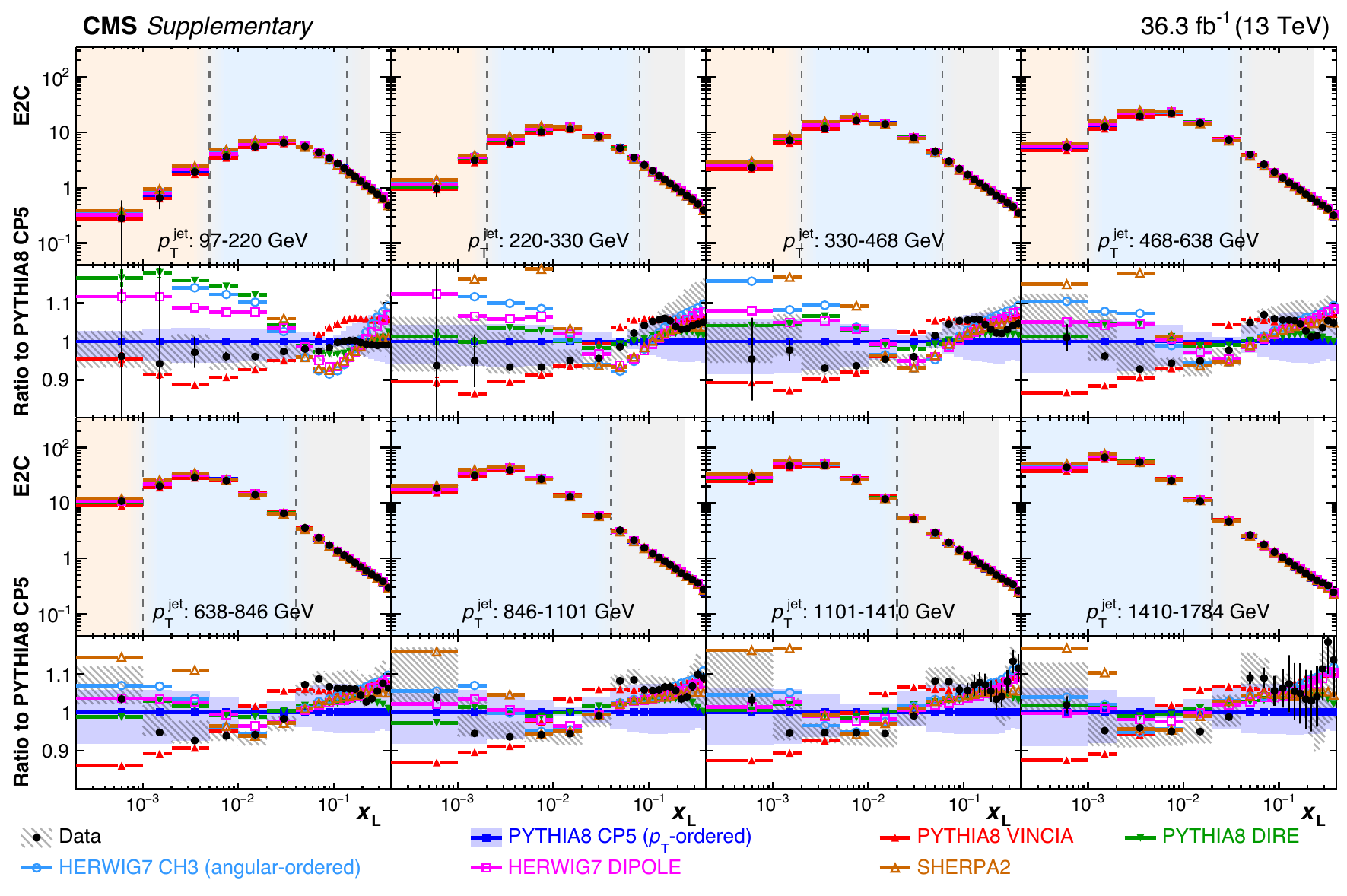}

\caption
{
Measured (unfolded) E2C distributions compared with the PYTHIA8 CP5 prediction (upper) and all the MC predictions (lower) in the eight \pt regions. Statistical and experimental systematic uncertainties are shown with error bars and hatched bands, respectively. The theory uncertainty in the PYTHIA8 CP5 prediction is shown with the blue bands. 
\label{fig:E2C_sup}
}
\end{figure*}

\begin{figure*}[htbp]
\centering
\includegraphics[width=\linewidth]{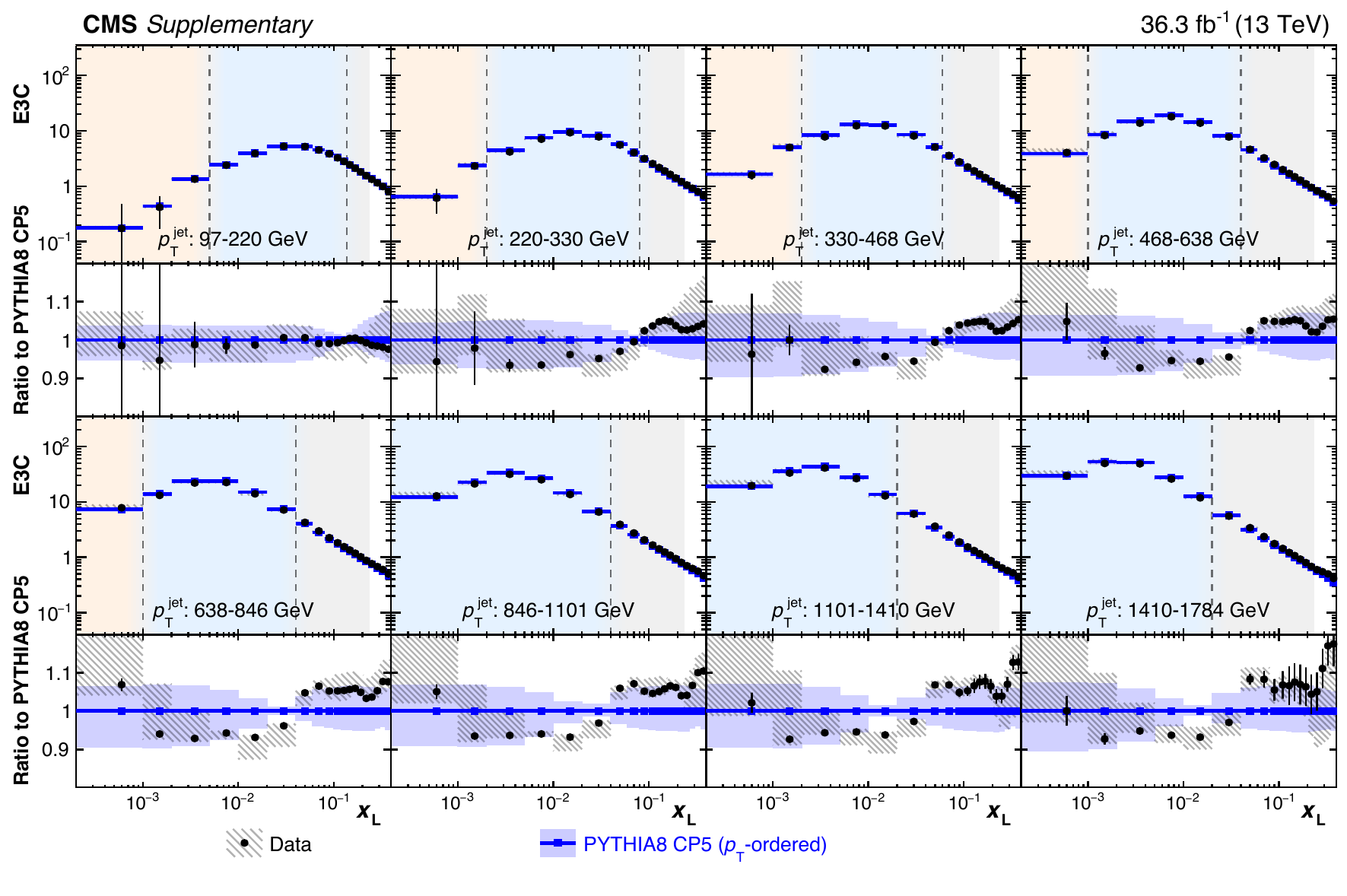}
\includegraphics[width=\linewidth]{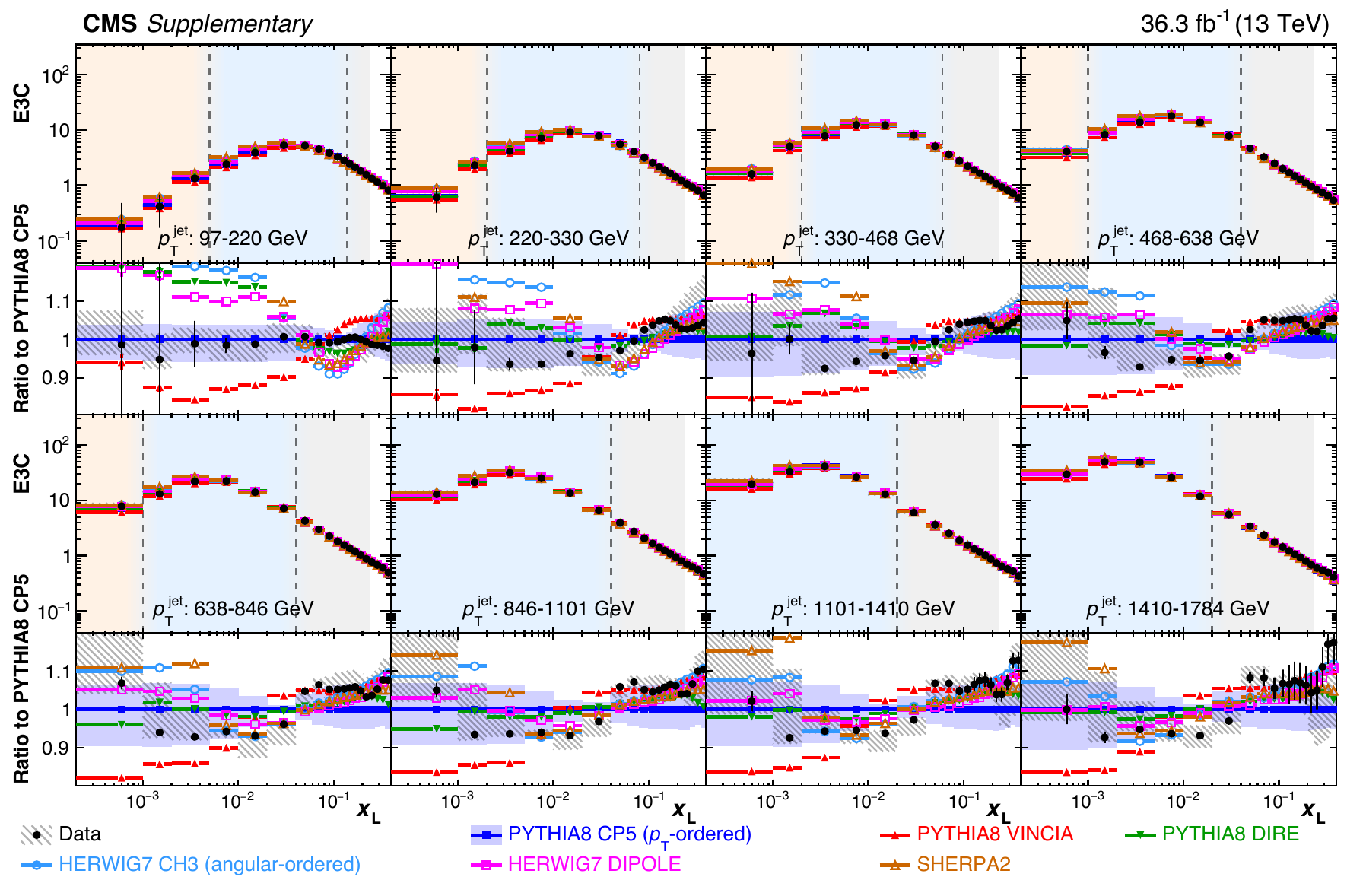}

\caption
{
Measured (unfolded) E3C distributions compared with the PYTHIA8 CP5 prediction (upper) and all the MC predictions (lower) in the eight \pt regions. Statistical and experimental systematic uncertainties are shown with error bars and hatched bands, respectively. The theory uncertainty in the PYTHIA8 CP5 prediction is shown with the blue bands. 
\label{fig:E3C_sup}
}
\end{figure*}

\begin{figure*}[htbp]
\centering
\includegraphics[width=\linewidth]{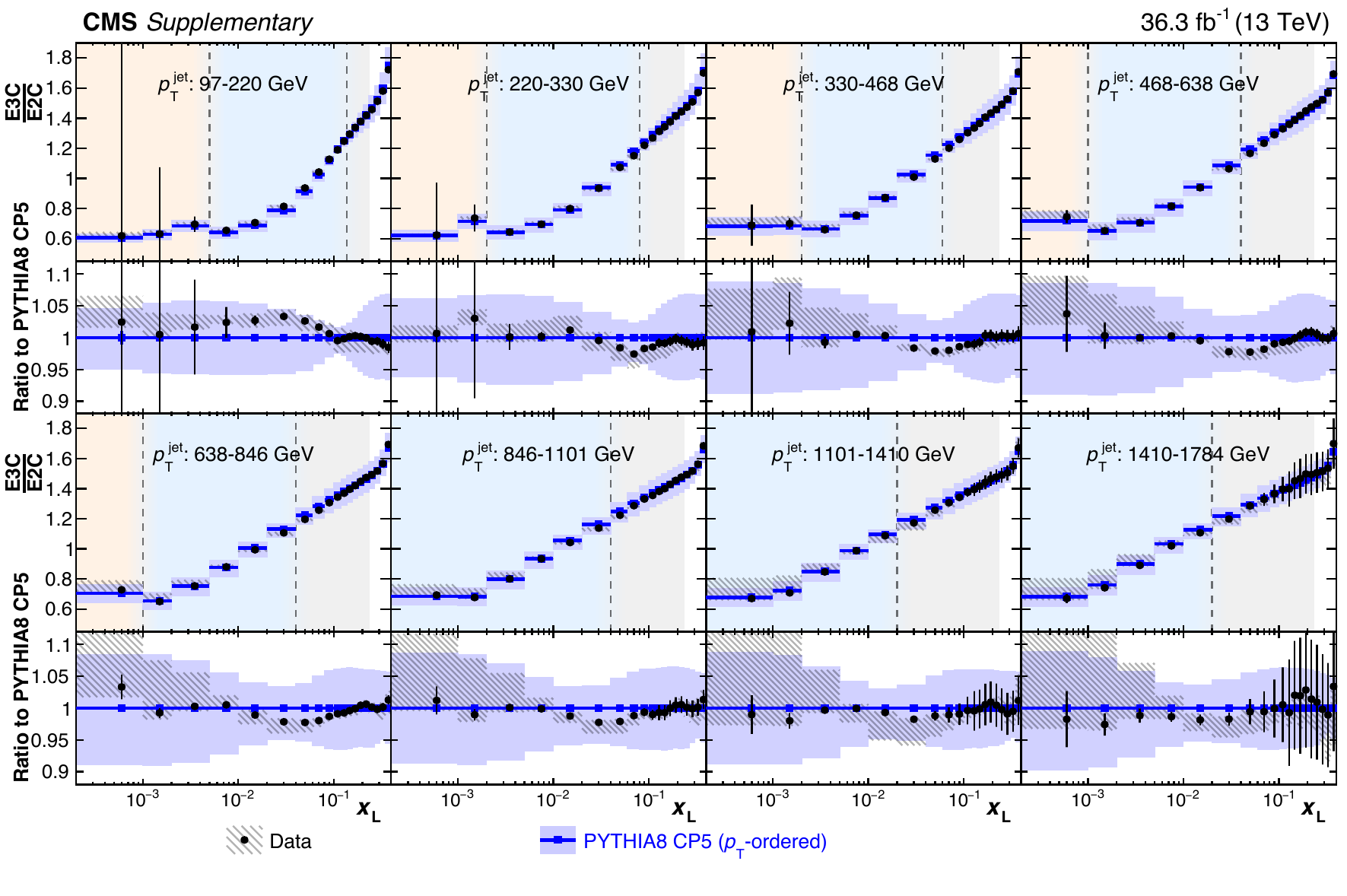}
\includegraphics[width=\linewidth]{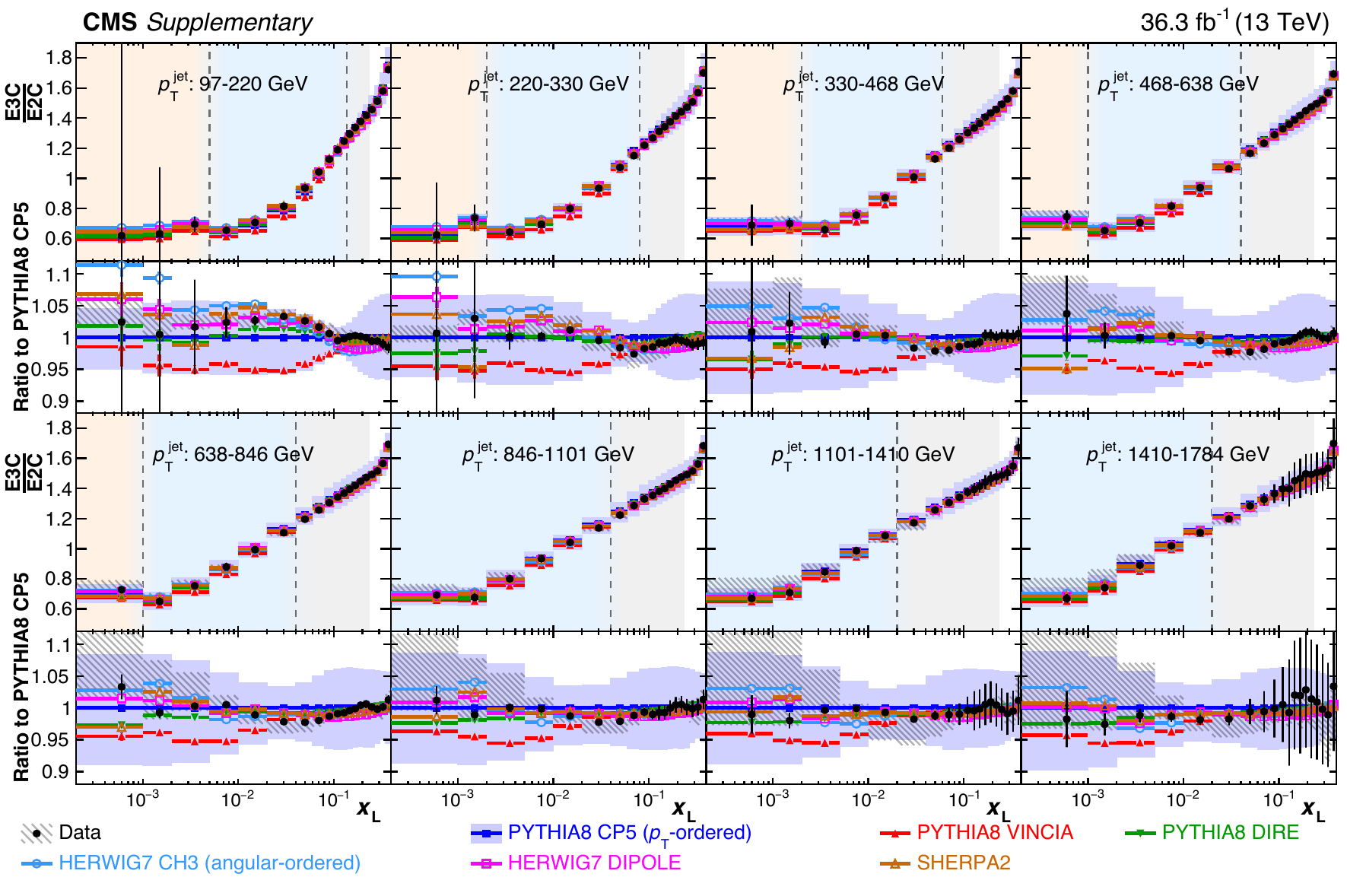}

\caption
{
Ratio of the unfolded E3C and E2C distributions compared with the PYTHIA8 CP5 prediction (upper) and all the MC predictions (lower) in the eight \pt regions. Statistical and experimental systematic uncertainties are shown with error bars and hatched bands, respectively. The theory uncertainty in the PYTHIA8 CP5 prediction is shown with the blue bands. 
\label{fig:ratio_sup}
}
\end{figure*}

\begin{figure*}[htbp]
\centering
\includegraphics[width=\linewidth]{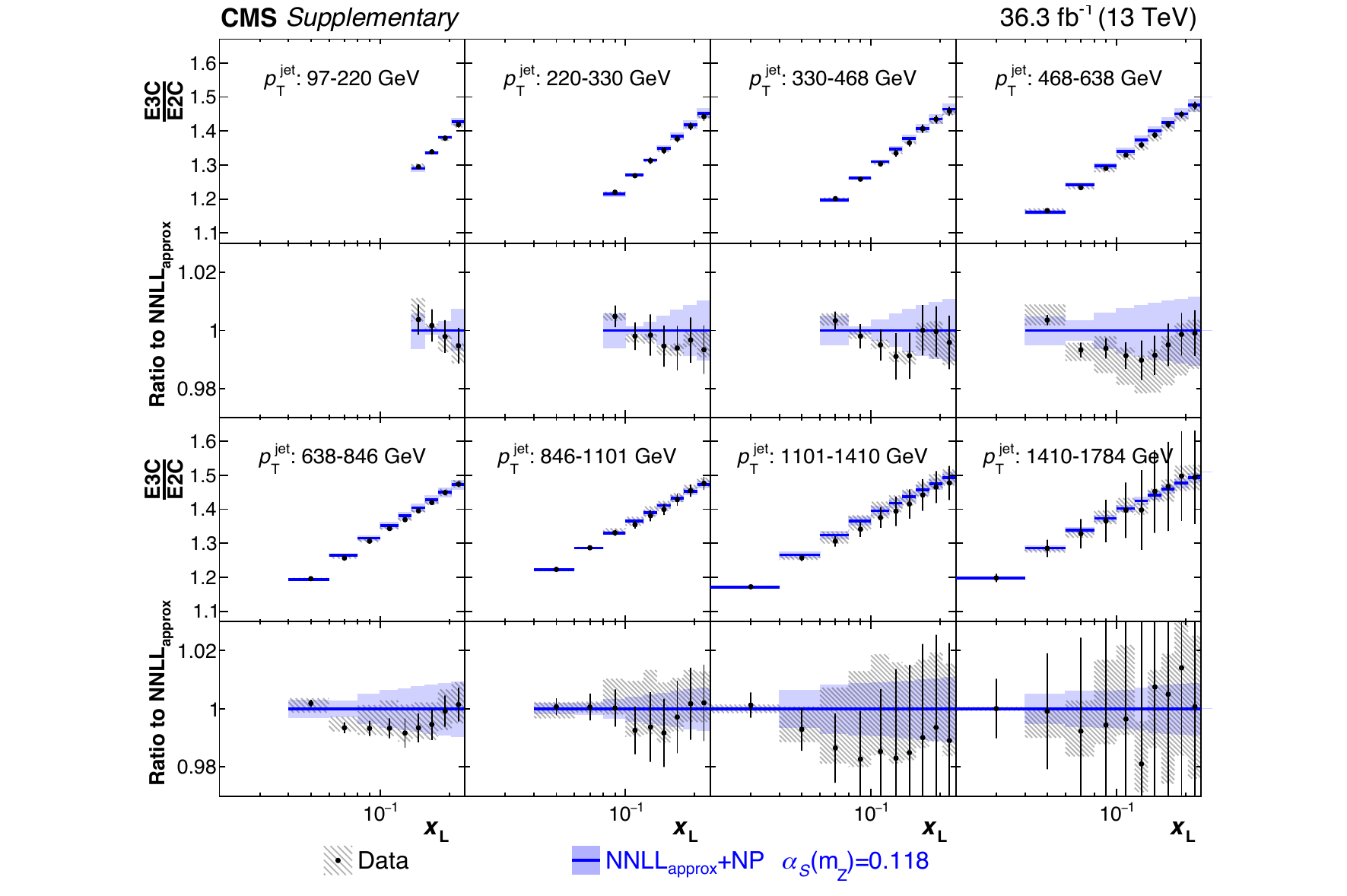}
\caption
{
Ratio of the unfolded E3C and E2C distributions in the eight \pt regions. The \NNLL theoretical predictions~\cite{Chen:2023zlx} and the corresponding uncertainties are corrected to hadron-level and normalized to the measured data. The statistical and experimental systematic uncertainties are shown with error bars and boxes, respectively.
\label{fig:ratio_theo_sup}
}
\end{figure*}

\begin{figure*}[htbp]
\centering
\includegraphics[width=0.8\linewidth]{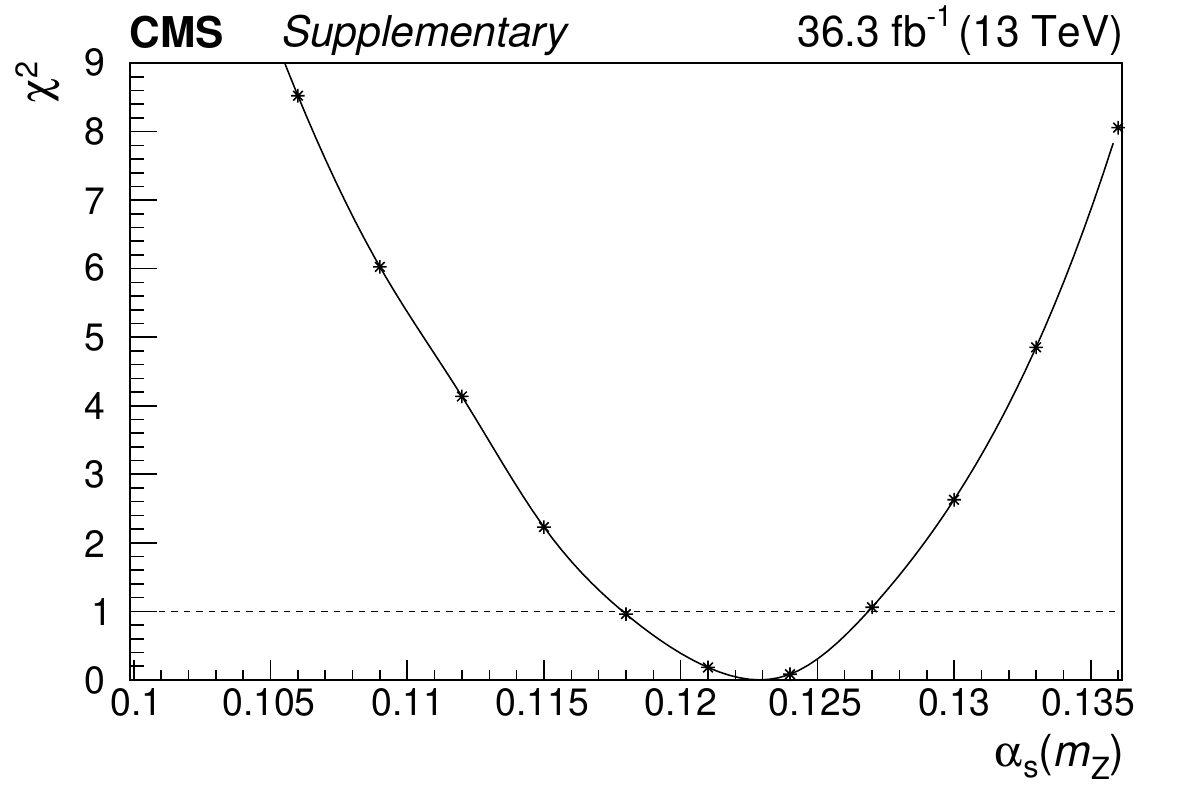}
\caption
{
The $\chi^2$ values determined from the fit of the measured data, as a function of \asZ.
\label{fig:chi2}
}
\end{figure*}

\begin{figure*}[htbp]
\centering
\includegraphics[width=0.8\linewidth]{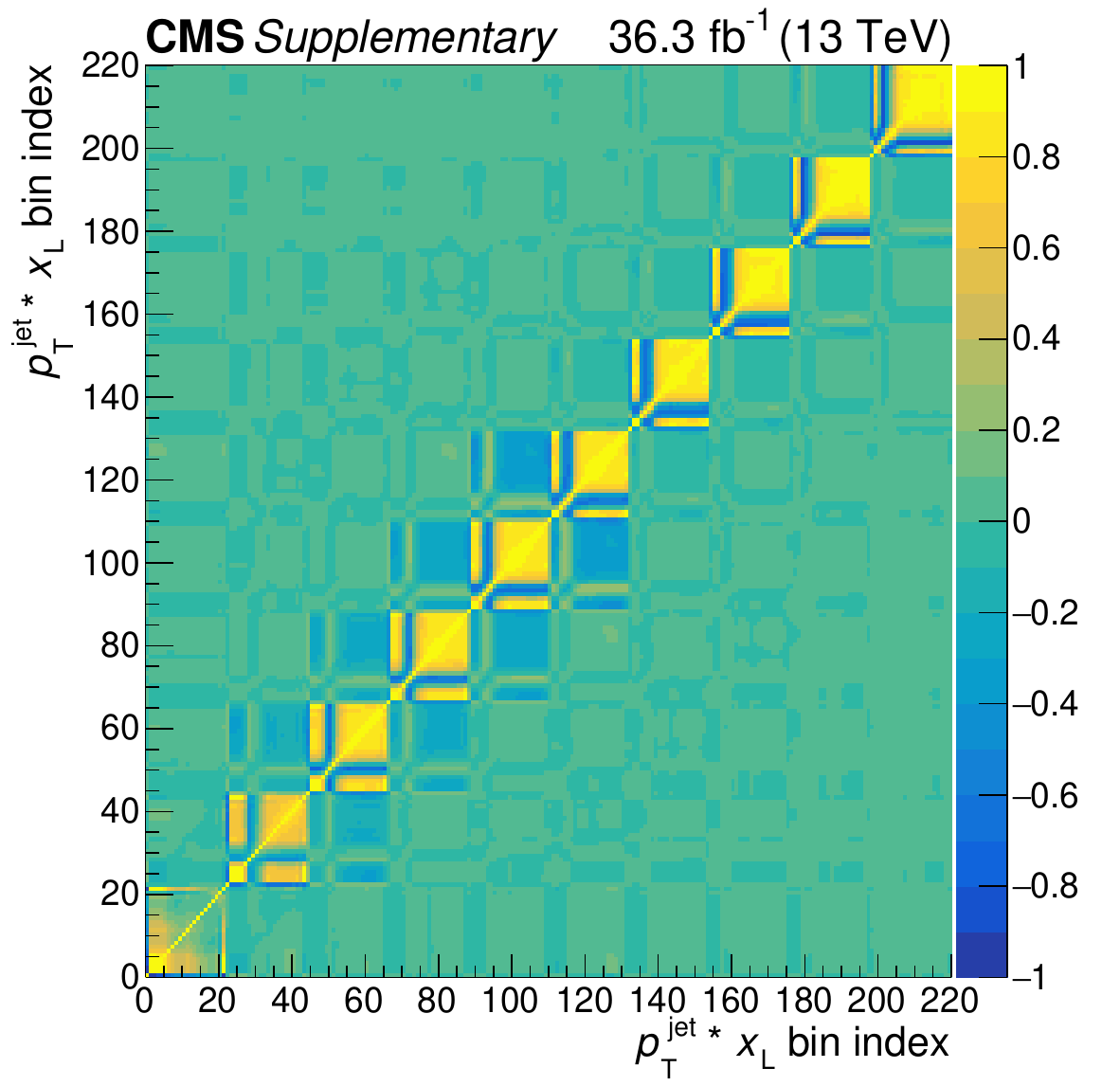}
\caption
{
The statistical correlation matrix of the bins of the unfolded E2C distribution. The bin number $n$ in this figure is defined by $n=22 i_{\pt}  + i_{\xL}$, where $i_{\pt}$ and $i_{\xL}$ are the indices of the \pt and \xL bins. In total there are ten \pt bins, the eight presented in the analysis plus the overflow and underflow bins. The number of \xL bins per \pt region is 22.
\label{fig:E2C_cor}
}
\end{figure*}
\begin{figure*}[htbp]
\centering
\includegraphics[width=0.8\linewidth]{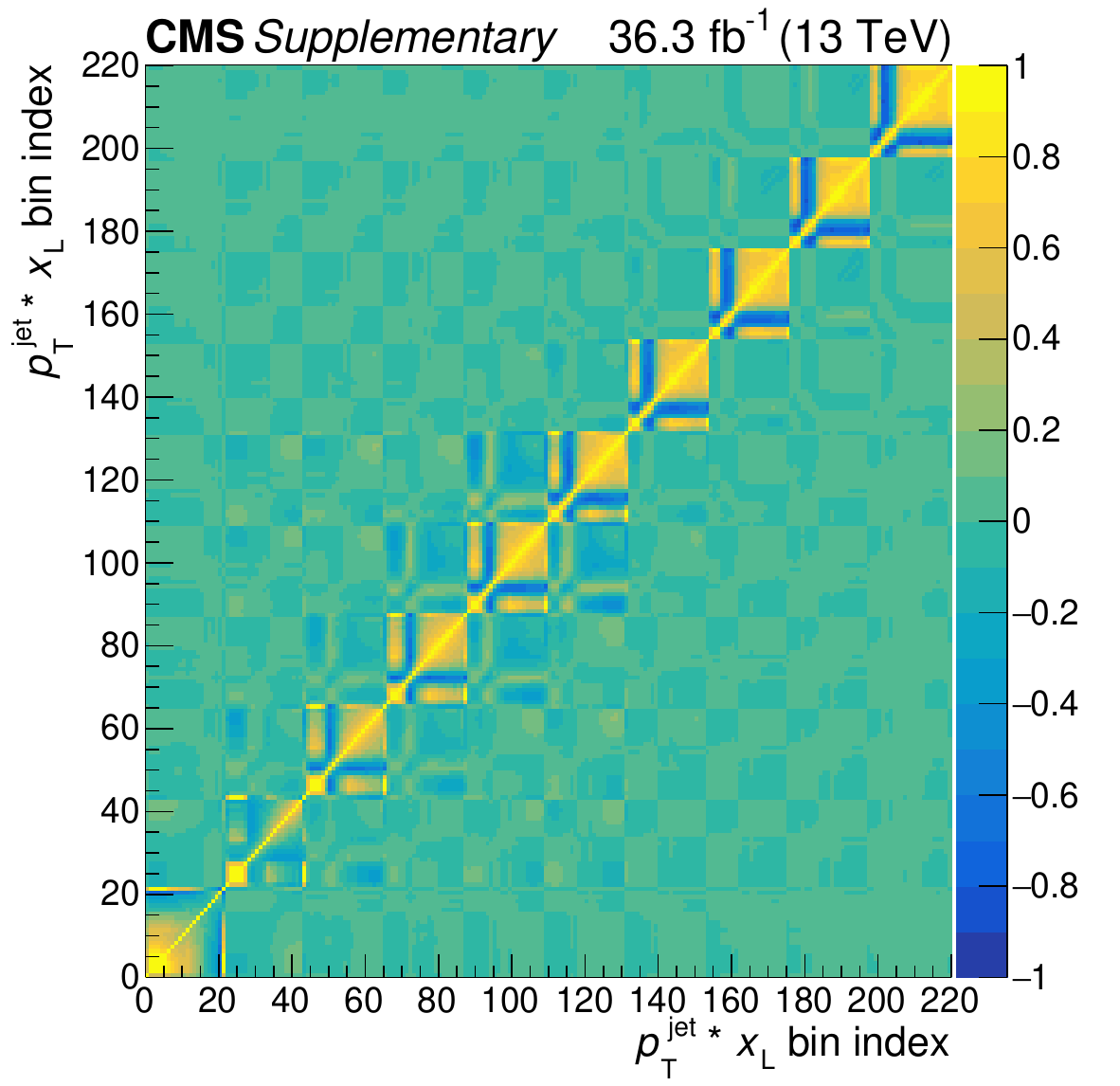}
\caption
{
The statistical correlation matrix of the bins of the unfolded E3C distribution. The bin number $n$ in this figure is defined by $n=22 i_{\pt}  + i_{\xL}$, where $i_{\pt}$ and $i_{\xL}$ are the indices of the \pt and \xL bins. In total there are ten \pt bins, the eight presented in the analysis plus the overflow and underflow bins. The number of \xL bins per \pt region is 22.
\label{fig:E3C_cor}
}
\end{figure*}
\begin{figure*}[htbp]
\centering
\includegraphics[width=0.8\linewidth]{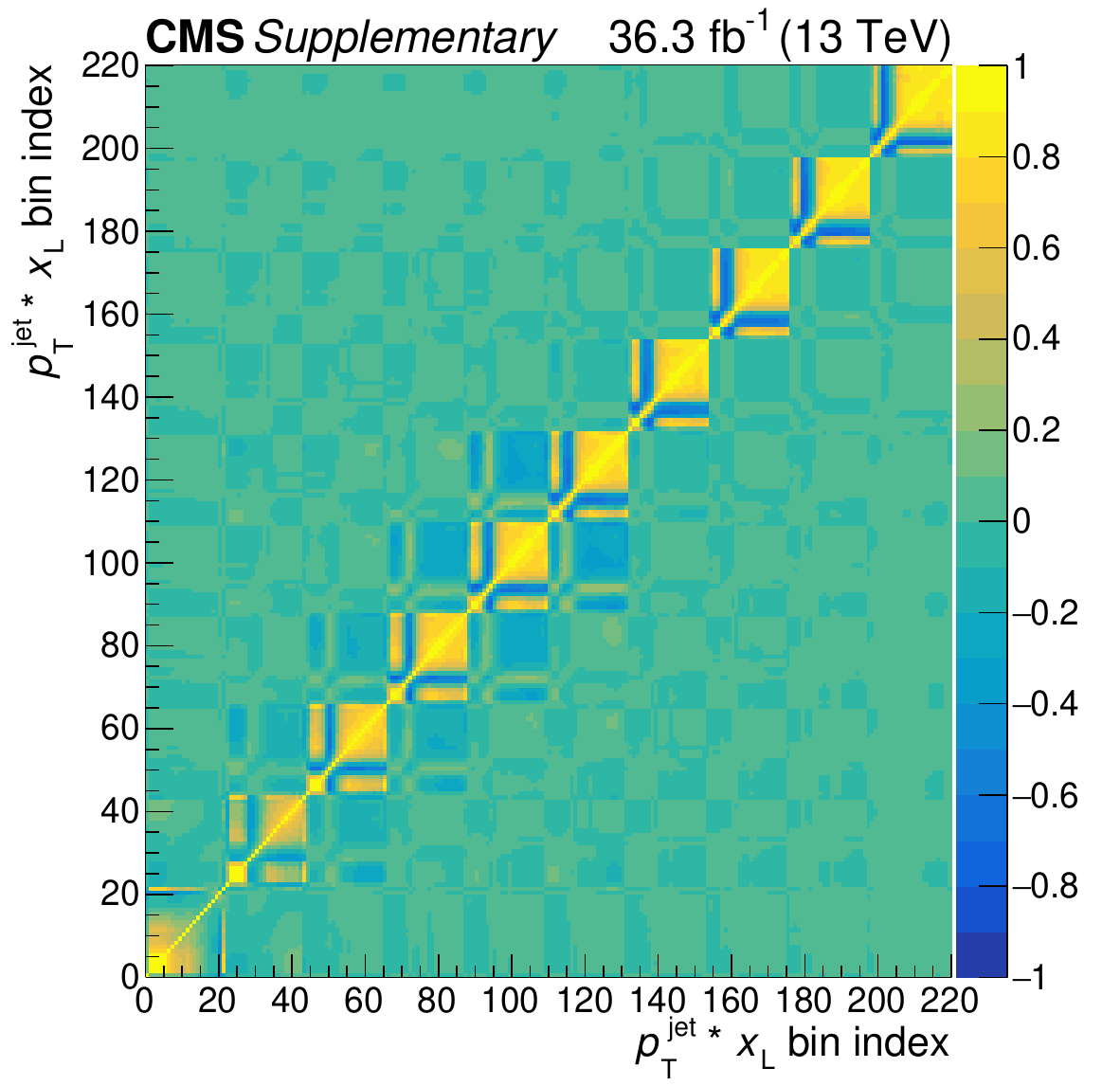}
\caption
{
The statistical correlation matrix of the bins of the unfolded E3C/E2C distribution. The bin number $n$ in this figure is defined by $n=22 i_{\pt}  + i_{\xL}$, where $i_{\pt}$ and $i_{\xL}$ are the indices of the \pt and \xL bins. In total there are ten \pt bins, the eight presented in the analysis plus the overflow and underflow bins. The number of \xL bins per \pt region is 22.
\label{fig:ratio_cor}
}
\end{figure*}

}
\cleardoublepage \section{The CMS Collaboration \label{app:collab}}\begin{sloppypar}\hyphenpenalty=5000\widowpenalty=500\clubpenalty=5000
\cmsinstitute{Yerevan Physics Institute, Yerevan, Armenia}
{\tolerance=6000
A.~Hayrapetyan, A.~Tumasyan\cmsAuthorMark{1}\cmsorcid{0009-0000-0684-6742}
\par}
\cmsinstitute{Institut f\"{u}r Hochenergiephysik, Vienna, Austria}
{\tolerance=6000
W.~Adam\cmsorcid{0000-0001-9099-4341}, J.W.~Andrejkovic, T.~Bergauer\cmsorcid{0000-0002-5786-0293}, S.~Chatterjee\cmsorcid{0000-0003-2660-0349}, K.~Damanakis\cmsorcid{0000-0001-5389-2872}, M.~Dragicevic\cmsorcid{0000-0003-1967-6783}, P.S.~Hussain\cmsorcid{0000-0002-4825-5278}, M.~Jeitler\cmsAuthorMark{2}\cmsorcid{0000-0002-5141-9560}, N.~Krammer\cmsorcid{0000-0002-0548-0985}, A.~Li\cmsorcid{0000-0002-4547-116X}, D.~Liko\cmsorcid{0000-0002-3380-473X}, I.~Mikulec\cmsorcid{0000-0003-0385-2746}, J.~Schieck\cmsAuthorMark{2}\cmsorcid{0000-0002-1058-8093}, R.~Sch\"{o}fbeck\cmsorcid{0000-0002-2332-8784}, D.~Schwarz\cmsorcid{0000-0002-3821-7331}, M.~Sonawane\cmsorcid{0000-0003-0510-7010}, S.~Templ\cmsorcid{0000-0003-3137-5692}, W.~Waltenberger\cmsorcid{0000-0002-6215-7228}, C.-E.~Wulz\cmsAuthorMark{2}\cmsorcid{0000-0001-9226-5812}
% ...
\par}
\vskip\cmsinstskip
\dag:~Deceased\\
$^{1}$Also at Yerevan State University, Yerevan, Armenia\\
$^{2}$Also at TU Wien, Vienna, Austria\\
% ...
\end{sloppypar}
\end{document}